\documentclass[a4paper,11pt]{article}
\pdfoutput=1 

\usepackage{jcappub} 

\usepackage[T1]{fontenc} 
\usepackage[utf8]{inputenc}
\usepackage{subcaption}

\title{Bounds on $\Lambda$ at the Galactic Center}

\author[a,b,c 1]{Prajwal Hassan Puttasiddappa,\note{Corresponding author.}}
\author[d]{Muzammil Mushtaq,}
\author[b,e,f]{William Ramirez}
\author[a]{and David F. Mota}

\affiliation[a]{Institute of Theoretical Astrophysics, University of Oslo, Norway}
\affiliation[b]{PPGCosmo, Universidade Federal do Esp\'irito Santo, Brazil}
\affiliation[c]{Departamento de F\'isica Te\'orica, Universidade do Estado do Rio de Janeiro, Brazil}
\affiliation[d]{Zentrum f\"ur Astronomie, Heidelberg University, Germany}
\affiliation[e]{Centro Brasileiro de Pesquisas F\'isicas, Brazil}
\affiliation[f]{Physik-Department, Technische Universit\"at M\"unchen, Germany}

\emailAdd{prajwal.puttasiddappa@edu.ufes.br}

\abstract{We constrain the cosmological constant, $\Lambda$, using astrometric and spectroscopic observations of the S2 star, supplemented by S1 and S14 stars orbiting Sgr A$^*$. We develop a framework for modelling relativistic stellar orbits in general spherically symmetric spacetimes and perform Bayesian parameter estimation to jointly infer orbital and spacetime parameters. Applying this framework to Schwarzschild-de Sitter spacetime, we obtain upper bounds on $\Lambda$. The individual constraints exhibit the expected dependence on the dynamical timescales of the stars, with the longer-period S1 providing the strongest sensitivity. Combining the individual constraints, we obtain $\Lambda \lesssim 6.9\times10^{-48}\mathrm{m}^{-2}$ at 68\% credibility and $\Lambda \lesssim 1.0\times10^{-38}\mathrm{m}^{-2}$ at 95\% credibility. The 68\% constraint is the strongest bound at a comparable dynamical timescale.}

\begin{document}
\maketitle
\flushbottom
\vspace{1cm}

\section{Introduction}

Standard cosmological probes, including Type Ia supernovae \cite{SupernovaSearchTeam1998fmf, SupernovaCosmologyProject1998vns}, Baryon Acoustic Oscillations \cite{BAO1, BAO2, BAo3}, and the Cosmic Microwave Background (CMB) \cite{Planck2018CMB}, have established the late-time accelerated expansion of the Universe, conventionally attributed to dark energy. Recent results from the Dark Energy Spectroscopic Instrument (DESI) have hinted at possible evolution of the dark energy equation of state \cite{DESI2, DESI1, DESI3DR2}, although the evidence remains inconclusive \cite{DESIDR2}. Within the standard $\Lambda$CDM framework, the cosmological constant $\Lambda$ provides the minimal and most successful phenomenological description of this acceleration. Although its effects are most apparent on cosmological scales, its manifestation in local, gravitationally bound systems remains a subject of interest.

The effects of dark energy on the Local Group and nearby galaxy clusters have been studied extensively \cite{Chernin2000, Baryshev2000kw, Karachentsev:2003eh, Teerikorpi2005zh, Chernin2009localgroup, Chernin2010DarkED, Chernin2015localflow, Benisty:2023vbz} (also see, \cite{Silbergleit2019oyx}). In a bound system, the repulsive linear contribution associated with $\Lambda$ competes with Newtonian attraction and defines a zero-velocity radius where the two effects balance. This provides a natural framework for describing the nearly linear and dynamically `quiet' local Hubble flow. A nonzero $\Lambda$ also modifies the timing argument used to infer the mass of systems such as the Milky Way-Andromeda pair \cite{galacticdynamicsbook, LiandWhite, Partridge2013dsa, McLeod2016EstimatingTM, Lemos:2020vhj, Benistytimingargument, Benisty2024galaxygroups}.

Existing constraints, summarized in Table~\ref{tab:Lambdabounds}, span many orders of magnitude, from Solar-System dynamics to cosmological observations. The hierarchy of these constraints is closely related to the characteristic dynamical timescale of the system \cite{Benisty:2023vbz}. For Schwarzschild-de Sitter (SdS) spacetime, the Kretschmann scalar can be written as \cite{Benisty2023clf}:

\begin{equation}\label{kretschmann}
    K_{\rm SdS} = K_{\rm Sch}\left[1 + \frac{1}{72}\left(\frac{T_K}{T_\Lambda}\right)^4\right]\ ,
\end{equation}
where $K_{\rm Sch}=48G^2M^2/(c^4 r^6)$ is the Kretschmann scalar for Schwarzschild spacetime, $T_K$ is the Keplerian dynamical timescale ($2\pi\sqrt{r^3/GM}$), and $T_\Lambda = 2\pi/(c\sqrt{\Lambda})$ is the characteristic timescale associated with $\Lambda$, which, for the observed cosmological value, $T_\Lambda\sim60$~Gyr, is much larger than the dynamical timescales of bound systems, explaining why local constraints are generally weaker than cosmological determinations.

The immediate environment of Sagittarius A$^*$ (Sgr A$^*$) at the Galactic Center (GC) presents two observational probes for testing gravity: its horizon-scale shadow and the relativistic orbits of surrounding S-stars. The Event Horizon Telescope (EHT) has resolved horizon-scale emissions of Messier $87^*$ \cite{EventHorizonTelescope2019dse}, and Sgr A$^*$ \cite{EventHorizonTelescope2022wkp}, finding evidence broadly consistent with the Kerr geometry, while current uncertainties in shadow-size measurements still allow departures from the standard black hole spacetime. The angular diameter of the black hole shadow, particularly in the case of Sgr A$^*$, which probes curvature scales higher than M$87^*$, remains a vital diagnostic to place constraints on the underlying spacetime structure \cite{Vagnozzi2022horizon}. 

Stellar orbits around Sgr A$^*$ provide a complementary dynamical probe of the Galactic Center, enabled by long-term monitoring with the Keck Observatory \cite{KeckGhez, KeckHees2017} and the Very Large Telescope (VLT) \cite{DataGillessen2016}, and further enhanced by the GRAVITY instrument \cite{GRAVITY2021Massdist}. These observations have established a population of young stars orbiting the central black hole; among which the S2 (or S0-2) star provides the most stringent constraints because of its short $\sim16$-year period and highly eccentric orbit. Near the periapsis, it reaches velocities near $3\%$ of the speed of light. Its orbit has enabled direct measurements of relativistic effects, including the gravitational redshift \cite{GRAVITY2018redshift, Keckredshift} and periapsis precession \cite{GRAVITY2020precession}. The astrometry and radial velocity data have been used to test a wide range of modified gravity \cite{scalartensorvector, scalartensorvectorMOG, Benisty:2021davis, fRS2, HorndeskiS2, beyondHorndeski} and nonstandard compact-object scenarios \cite{unveilingwormhole, nakedsingularity, testingbhspacetimebayesian, Bambhaniyatesting, RARold, RARrecent, ZakharovRAR}. Furthermore, the S2 orbit has been utilized to constrain dark matter distributions \cite{narrowbosonDM, ultralightbosnicDM, DMspike}.

\begin{table*}
\centering
\begin{tabular}{l c}
\hline
\hline
Observed effect & Estimate on $\Lambda \ (\mathrm{m^{-2}})$ \\
\hline
\multicolumn{2}{c}{\textit{Solar System estimates}}\\
\hline
Light deflection & No constraint \\
Gravitational time delay \cite{Kagramanova:2006ax} & $ \lesssim 10^{-24}$ \\
Geodetic precession \cite{Kagramanova:2006ax} & $\lesssim 10^{-27}$ \\
Gravitational redshift \cite{Kagramanova:2006ax} & $ \lesssim 10^{-28}$ \\
Mercury Perihelion shift \cite{Kagramanova:2006ax} & $ \lesssim 10^{-41}$ \\
Perihelion precessions of Earth and Mars \cite{Sereno:2006re} & $\lesssim 10^{-42}$ \\
Perihelion precession of Saturn \cite{Benisty2023clf} & $\lesssim 10^{-46}$ \\
\hline
\multicolumn{2}{c}{\textit{Astrophysical / Cosmological estimates}} \\
\hline
Binary pulsars \cite{Jetzer:2006gn} & $\lesssim 10^{-40}$ \\
Lensing by Galaxy cluster \cite{Ishak:2007ea} & $\lesssim 10^{-50}$ \\
Galaxy rotation curves \cite{Benistyrotationcurve} & $ \lesssim 1.9 \times 10^{-51}$ \\
Local group \cite{Benisty2023clf} & $= (3.13 \pm 2.42) \times  10^{-52}$ \\
Virgo cluster \cite{Benisty2023clf} & $> 1.58 \times  10^{-53}$ \\
LSS/CMB \cite{Planck2018CMB} & $= (1.097 \pm 0.02)\times10^{-52}$ \\
SH0ES \cite{Riess2021SH0ES} & $ = (1.29 \pm 0.05) \times 10^{-52}$ \\
Supernovae Ia (Pantheon+) \cite{Pantheonplus} & $ \approx 1.1 \times 10^{-52}$ \\
\hline
\hline
\end{tabular}
\caption{Existing bounds and inferred values of $\Lambda$ from local, astrophysical, and cosmological observations.}
\label{tab:Lambdabounds}
\end{table*}

Despite probing a stronger gravitational potential, S2's dynamical timescale is comparable to that of the outer Solar System planets. This makes the GC particularly useful for testing whether the timescale hierarchy seen in existing $\Lambda$ constraints extends into the strong-potential regime. The combined dark-energy and post-Newtonian effects on the S2 orbit were studied in \cite{Benisty2023clf}, where $\Lambda$ was constrained from its relativistic periapsis precession. Here we instead use the full astrometric and spectroscopic information to constrain $\Lambda$ by modelling the relativistic stellar trajectories. We use S2 as the primary probe and supplement it with S1 and S14. Our analysis employs \texttt{SNOPy}, a Python framework we developed to numerically integrate relativistic timelike geodesics in spherically symmetric spacetimes and jointly infer the spacetime, orbital, and reference-frame parameters from the observations.

The remainder of this paper is organized as follows. In Section~\ref{sec:firstbounds}, we establish the basic formalism and derive bounds on $\Lambda$ from the shadow size of Sgr A$^*$ and S2 precession (derivations in Appendix~\ref{app:shadow}). Section~\ref{sec:dataandanalysis} describes the orbit modelling framework, observational data, and statistical analysis (with implementation details in Appendix~\ref{app:orbitmodel}). In Section~\ref{sec:results}, we present the results and discuss their implications, followed by conclusions in Section~\ref{sec:conclusion}.

\section{Constraints on $\Lambda$ from the Sgr A$^*$ shadow and S2 precession}\label{sec:firstbounds}

We model the stellar dynamics and light propagation in the Schwarzschild-de Sitter (SdS) or the Kottler \cite{Kottler1918cxc} spacetime. The line element in spherical coordinates is given by, 

\begin{equation}\label{sdsmetric}
\begin{split}
    ds^2 &= -f(r) dt^2 + \frac{1}{f(r)} dr^2 + r^2 ( d\theta^2 + \sin^2\theta d\varphi^2)\ ,\\
    \text{ where , }\quad f(r) &= \left(1 - \frac{2GM}{rc^2} - \frac{\Lambda r^2}{3}\right)\ .
\end{split} 
\end{equation}
which reduces to the Schwarzschild solution in the limit $\Lambda = 0$. For non-vanishing $\Lambda$, the spacetime is no longer asymptotically flat, and global quantities such as the Arnowitt-Deser-Misner (ADM) mass are not well-defined. Nevertheless, on the dynamical scales relevant to the S-stars and the photon trajectories, the parameter $M$ remains an accurate characterization of the central gravitating mass. Throughout this work, we assume a positive cosmological constant\footnote{For a discussion on $\Lambda < 0$, see \cite{sdsasdsshadow1, ZakharovAdS, Prado-Fuentes2025nvl}} ($\Lambda > 0$), consistent with the SdS structure. For $9\Lambda(GM/c^2)^2 < 1$, $f(r) = 0$ admits two roots corresponding to a black hole horizon and a cosmological horizon, with the orbits of interest confined to the static region between them. 

To describe null ($\epsilon = 0$) and timelike ($\epsilon = 1$) trajectories, we consider geodesics satisfying $g_{\mu\nu}\dot{x}^\mu \dot{x}^\nu = -\epsilon c^2$, where dots denote differentiation with respect to an affine parameter $\lambda$ which is taken as the proper time $\tau$ when $\epsilon = 1$. In terms of the conserved specific energy $E = f(r)\dot{t}$ and specific angular momentum $L = r^2\dot{\varphi}$, the radial equation of motion becomes:

\begin{equation}\label{radialnormalization}
\begin{split}
    \dot{r}^2 &= E^2 - V(r)\ ,\\
    \text{with , }\quad V(r) &= \left(1 - \frac{2GM}{rc^2} - \frac{\Lambda r^2}{3}\right) \left(\epsilon c^2 + \frac{L^2}{r^2}\right)\ .
\end{split}
\end{equation}
Substituting $u = 1/r$ and differentiating with respect to $\varphi$ gives

\begin{equation}\label{secondordgeo}
    \frac{d^2u}{d\varphi^2} + u = \frac{3GM}{c^2}u^2 + \epsilon\left(\frac{GM}{L^2} - \frac{\Lambda c^2}{3L^2 u^3}\right).
\end{equation}

Before performing full orbital fits, we examine single-measurement constraints on $\Lambda$ from the angular size of the Sgr A$^*$ shadow and from the periapsis precession of S2.

\subsection{Sgr A$^*$ shadow}\label{sec:shadow}

For null geodesics ($\epsilon = 0$), the equation \eqref{secondordgeo} is independent of $\Lambda$ \cite{rindlerbook}, indicating that the effect of $\Lambda$ on null trajectories arises not through the geodesic dynamics, but through the geometry of the spacetime. Also the equation \eqref{radialnormalization} simplifies in terms of the impact parameter $b(r) = L/E$ to, $\dot{r}^2/L^2 = 1/b^2 - f(r)/r^2$. The photon-sphere radius follows from the extremum of $V_\gamma \equiv f(r)/r^2$, giving $r_\gamma = 3GM/c^2$, which again, is independent of $\Lambda$ \cite{sdsasdsshadow1}. The apparent boundary of the shadow is defined by null geodesics that asymptotically approach $r_\gamma$ and is determined by the critical impact parameter $b_{\text{crit}} \equiv b(r_\gamma)$, which separates captured and escaping trajectories. It is given by,

\begin{equation}\label{eq:bcrit}
    b_{\rm{crit}} = \frac{r_{\gamma}}{\sqrt{f(r_{\gamma})}} = \frac{3\sqrt{3}(GM/c^2)}{\sqrt{1 - 9\Lambda (GM/c^2)^2}}\ .
\end{equation}

Unlike asymptotically flat spacetimes where $b_{\rm{crit}}$ directly sets the shadow radius at infinity, the asymptotically de Sitter nature of SdS renders the standard impact parameter ill-defined. Consequently, the observed angular shadow size depends explicitly on the observer’s position and motion \cite{Firouzjaee:2019aij, Roy:2020dyy, Perlick:2018iye}. For a static observer at radial coordinate $r_O$ between the event and cosmological horizons, the angular size $\vartheta_{\rm sh}$ is given by \cite{Perlick:2018iye, Perlick:2021aok}, $\sin^2\vartheta_{\text{sh}} = b_{\text{crit}}^2 f(r_O)/r_O^2$. In the small-angle limit, the shadow radius is given by\footnote{See Appendix~\ref{app:shadow} for a brief discussion.} \cite{Vagnozzi2022horizon},

\begin{equation}\label{shadowrad}
    r_{\rm sh} = \frac{3\sqrt{3}(GM/c^2)}{\sqrt{1 - 9\Lambda (GM/c^2)^2}}\sqrt{1 - \frac{2GM}{r_O c^2} - \frac{\Lambda r_O^2}{3}}\ .
\end{equation}

Equation~\eqref{shadowrad} reduces to $3\sqrt{3}GM/c^2$ when taking the flat-space limit ($\Lambda \to 0$) prior to $r_O \to \infty$. Unlike Schwarzschild spacetime, the apparent shadow size depends explicitly on the observer’s position $r_O$, contracting to a point as $r_O$ approaches the cosmological horizon \cite{sdsshadownotes}. To compare with observations, the shadow radius is parameterized via the fractional deviation $\delta$ as $r_{\rm sh} = 3\sqrt{3} GM(\delta + 1)/c^2$. Because EHT measures angular size, bounds on $r_{\rm sh}$ (and thus $\Lambda$) depend directly on the mass-to-distance ratio ($M/D$), making them sensitive to $M$ and $D$ priors. To calibrate this scale, the average of dynamical estimates from Keck \cite{Keckredshift} and GRAVITY \cite{GRAVITY2020precession} is used to translate angular observations into the $r_{\rm sh}/M$ constraints (in units of $c^2/G$) shown in Figure~\ref{fig:sds_shadow} \cite{EventHorizonTelescope2022wkp}, implying $\Lambda \lesssim 10^{-41}$~m$^{-2}$ at $r_O \sim 8$~kpc \cite{Vagnozzi2022horizon}.

\subsection{S2 periapsis precession}\label{sec:precession}

For timelike geodesics $(\epsilon = 1)$, treating Schwarzschild and $\Lambda$ terms of \eqref{secondordgeo} as perturbations around the Newtonian solution $u_N = \frac{GM}{L^2}(1 + e\cos\varphi)$, one obtains the periapsis advance per orbit \cite{Kerrsdsprecession, Iorio:2007ub, Ruggiero:2010yn, Arakida:2012ya} (also in Appendix~\ref{app:shadow}),

\begin{equation}\label{eq:sdsprecession}
    \Delta\varphi \equiv \Delta\varphi_{\rm Schwarz} + \Delta\varphi_\Lambda = \frac{6\pi GM}{c^2 a(1-e^2)} + \frac{\Lambda \pi c^2 a^3}{GM}\sqrt{1-e^2}\ .
\end{equation}

A positive $\Lambda$ induces a small advance to the Schwarzschild precession. To quantify departures from the purely Schwarzschild prediction ($f_{\rm SP} = 1$), we define the dimensionless parameter $f_{\rm SP}$:

\begin{equation}
    f_{\rm SP} = 1 + \frac{\Lambda c^4 a^4 (1-e^2)^{3/2}}{6 G^2 M^2}\ .
\end{equation}

\begin{figure}[htbp]
    \centering
    \begin{subfigure}[b]{0.48\textwidth}
        \centering
        \includegraphics[width=\linewidth]{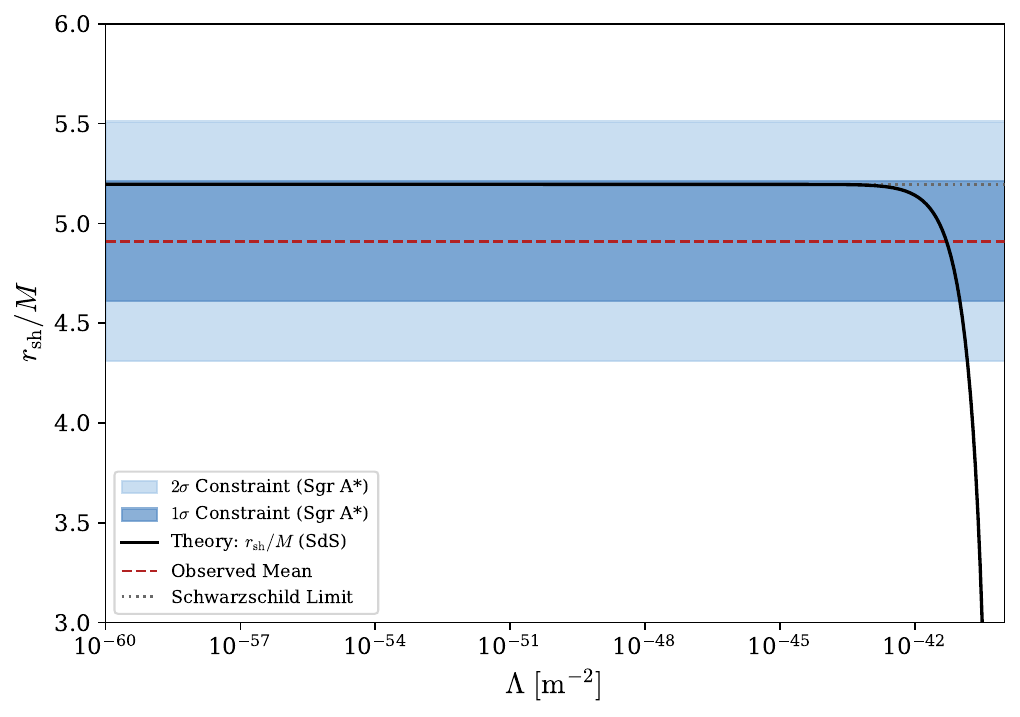}
        \caption{Shadow radius $r_{\rm sh}/M$ vs.\ $\Lambda$ for a static observer, with $1\sigma$ and $2\sigma$ Sgr A$^*$ constraints.}
        \label{fig:sds_shadow}
    \end{subfigure}
    \hfill
    \begin{subfigure}[b]{0.48\textwidth}
        \centering
        \includegraphics[width=\linewidth]{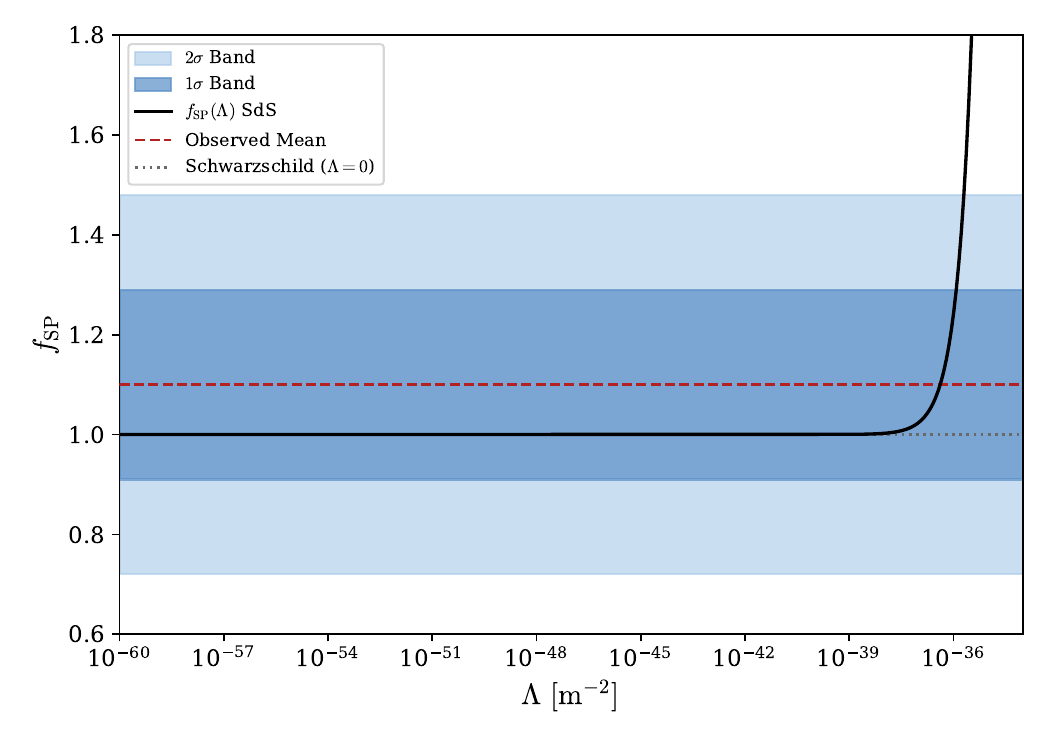}
        \caption{Precession parameter $f_{\rm SP}$ vs.\ $\Lambda$, with GRAVITY observational bounds \protect\cite{GRAVITY2020precession}.}
        \label{fig:sdsprecession}
    \end{subfigure}
    \caption{Observational constraints on $\Lambda$ from Sgr A$^*$ shadow size (left) and S2 periapsis precession (right).}
    \label{fig:combined_constraints}
\end{figure}

Comparing this result with the GRAVITY measurements $f_{\rm SP} = 1.10 \pm 0.19$ \cite{GRAVITY2020precession}, (Figure~\ref{fig:sdsprecession}), the current precision of the S2 precession measurement translates to an upper limit of $\Lambda \lesssim 10^{-36}$m$^{-2}$. A more stringent constraint of $\Lambda \lesssim 10^{-46}\text{ m}^{-2}$ was obtained in \cite{Benisty2023clf} utilizing a joint post-Newtonian and dark energy framework.

\section{Data, Modelling and Analysis}\label{sec:dataandanalysis}

\subsection{Orbit Modelling}

To model the stellar trajectories, we make use of \texttt{SNOPy}\footnote{\url{https://github.com/PrajwalPrem/SNOPy}}, a Python framework we developed for orbit integration and parameter inference in general spherically symmetric spacetimes, without relying on a post-Newtonian expansion. The geodesic equations

\begin{equation}\label{geodesic}
    \frac{d^2x^\mu}{d\tau^2} + \Gamma^\mu_{\alpha\beta} \frac{dx^\alpha}{d\tau}\frac{dx^\beta}{d\tau} = 0\ ,
\end{equation}
are integrated numerically using the SdS metric \eqref{sdsmetric}. The orbital parameters $(a,e)$ determine the pericenter $r_p=a(1-e)$ and apocenter $r_a=a(1+e)$, from which the conserved specific energy and angular momentum are determined by the turning-point conditions from \eqref{radialnormalization}. The orbit is then integrated forward and backwards in proper time to obtain the complete relativistic trajectory.

The trajectory in the orbital plane is subsequently transformed to the observer's frame using the orbital orientation angles $(i,\Omega,\omega)$ and the distance $D$ to Sgr A$^*$, yielding predicted sky-plane coordinates and line-of-sight velocities for direct comparison with astrometric and spectroscopic data. The motion and position of the astrometric reference frame are included in the modelling through the offset parameters. Relevant relativistic corrections to the observed signal, including the R{\o}mer time delay and relativistic redshift corrections, are taken into account. A detailed description of the orbit modelling implemented in \texttt{SNOPy} is discussed in Appendix~\ref{app:orbitmodel}.

The stellar orbit model is then described by the parameter vector,

\begin{equation}
    \Theta = \{ M, D, a, e, i, \Omega, \omega, t_{p}, x_0, y_0, v_{x0},  v_{y0}, v_{z0}, \Lambda \}\ ,
\end{equation}
where $M$ and $\Lambda$ characterize the spacetime, $D$ is the distance to the Sgr A$^*$, $(a,e,i,\Omega,\omega,t_p)$ are the orbital parameters, and $(x_0,y_0,v_{x0},v_{y0},v_{z0})$ describe the position and motion of the reference frame.

\subsection{Data}

We use astrometric and spectroscopic observations of S-stars from \cite{DataGillessen2016} and electronically accessible through the catalogue \cite{catalogue}. From the 17 stars in the dataset, S2 is our primary probe, supplemented by S1 and S14. Astrometric coordinates (RA, Dec) are reported in a reference frame centered approximately on the radio position of Sgr A$^*$, which carries an origin uncertainty of $\pm(0.2, 0.2)$~mas at the 2009.02 reference epoch \cite{Plewaoffset}. To account for residual calibration uncertainties between the astrometric frame and the dynamical center, we include nuisance parameters describing positional offsets and linear drifts in the orbital modelling introduced in the previous section.

For S2, we use 145 astrometric measurements and 44 radial-velocity measurements, covering the intervals 1992.224-2016.530 and 2000.487--2016.519, respectively \cite{DataGillessen2016}. The astrometry combines pre-2002 speckle observations with the SHARP camera at the ESO New Technology Telescope \cite{NTT} and post-2002 adaptive-optics observations with NACO at the Very Large Telescope \cite{NACO,NACO2}. The radial-velocity data were obtained with NIRC2 at Keck before 2003 and with Spectrograph for INtegral Field Observations in the Near Infrared (SINFONI) at the VLT from 2003 onward \cite{velocity1, velocity2, velocity3}.

For S1, we use 161 astrometric and 29 radial-velocity measurements, covering 1992.224-2016.530 and 2003.271-2016.519, respectively. Although stars such as S24 and S54 have longer orbital periods ($\gtrsim166$ yr), we retain S1 because its available observations cover a comparatively larger fraction of its orbit. For S14, we use 99 astrometric and 12 radial-velocity measurements, spanning 1992.224-2016.530 and 2004.537-2016.284, respectively. It has one of the highest eccentricities in the sample $(e\sim 0.97)$.
    
\subsection{Analysis}

The observational data vector $\mathcal{D}$ consists of the measured astrometric positions $(\mathrm{RA}_j,\mathrm{Dec}_j)$ and radial velocities $v_j^R$ at observation times $t_j^{\rm obs}$. For a given parameter set $\Theta$, the relativistic orbit model generates the stellar trajectory and its corresponding astrometric and spectroscopic observables. Because numerical integration points do not generally coincide with observational epochs $t_j^{\rm obs}$, we apply cubic-spline interpolation to the integrated trajectory. At each $t_j^{\rm obs}$, the corresponding emission time is determined iteratively via the R{\o}mer-delay relation (see \eqref{romerdelay}). This yields the theoretical data vector $\mathcal{D}^{\rm th}(\Theta) = \{(\text{RA}_j^{\rm th}, \text{Dec}_j^{\rm th}, v_{R,j}^{\rm th})\}$, evaluated at the exact timestamps of the observations. Comparison between $\mathcal{D}$ and $\mathcal{D}^{\rm th}(\Theta)$ yields the astrometric and spectroscopic residuals used in likelihood analysis. The likelihood can be given as,  

\begin{equation}
\ln \mathcal{L}(\Theta) = -\frac{1}{2} \left[ \chi^2_{\text{pos}}(\Theta) + \chi^2_{\text{vel}}(\Theta) \right]\ ,
\end{equation}
where the $\chi^2$ terms follow the standard definition:
\begin{equation}
\chi^2_{\text{pos}, \text{vel}} = \sum_{j} \left( \frac{\mathcal{D}_j - \mathcal{D}_j^{\text{th}}(\boldsymbol{\Theta})}{\sigma_j} \right)^2\ .
\end{equation}

Because the relativistic geodesics are integrated directly, the periapsis precession and other relativistic effects are already included in $\mathcal{D}^{\rm th}(\Theta)$, with no additional precession term introduced in the likelihood. This is appropriate for the dataset used here \cite{DataGillessen2016}, which predates the observational detection of Schwarzschild precession \cite{GRAVITY2020precession}. 

We perform independent Bayesian parameter estimation for S2, S1, and S14 using the same 14-dimensional parameter vector $\Theta$. For all three stars, we adopt a broad logarithmic prior on the cosmological
constant,
\begin{equation}\label{lambdaprior}
    \log_{10}(\Lambda/\text{m}^{-2}) \in [-70, -20]\ ,
\end{equation}
allowing the sampler to explore several orders of magnitude without imposing a preferred physical scale. For S2, the remaining parameters are assigned flat priors, similar to those adopted in \cite{scalartensorvector,fRS2,testingbhspacetimebayesian}, as summarized in Table~\ref{tab:flatS2prior}.


\begin{table}
\centering
\begin{tabular}{lc}
\hline
\textbf{Parameter} & \textbf{Uniform Prior}\\
\hline
$M$ ($10^6\,M_\odot$) & [3, 5] \\
$D$ (kpc) & [7, 9]\\
$a$ (mas) & [110, 140]\\
$e$ & [0.85, 0.9]\\
$i$ ($^\circ$) & [130, 138]\\
$\Omega$ ($^\circ$) & [223, 231]\\
$\omega$ ($^\circ$) & [60, 70]\\
$t_p - 2002$ (yr) & [0.2, 0.4] \\
$x_0$ (mas) & [-10, 10]\\
$y_0$ (mas) & [-10, 10]\\
$v_{x0}$ (mas yr$^{-1}$) & [-0.1, 0.1]\\
$v_{y0}$ (mas yr$^{-1}$) & [-0.1, 0.1]\\
$v_{z0}$ (km s$^{-1}$) & [-50, 50]\\
\hline
\end{tabular}
\caption{Flat priors for S2 orbital parameter inference.}
\label{tab:flatS2prior}
\end{table}

For S1 and S14, we adopt a hybrid prior strategy based on the orbital constraints reported in \cite{DataGillessen2016}. Since all three stars are assumed to orbit the same central object, Gaussian priors on $M$ and $D$ are obtained from the posterior constraints of the S2 analysis (Table~\ref{tab:s2params}). The orbital elements are assigned Gaussian priors centered on the corresponding literature values from \cite{DataGillessen2016}. Gaussian priors on the reference-frame offsets are motivated by \cite{Plewaoffset}, while the prior on the line-of-sight velocity offset $v_{z0}$ follows \cite{DataGillessen2016}. The adopted priors are summarized in Table~\ref{tab:otherstars}. S1 and S14 are therefore used as supplementary probes to assess the consistency of the $\Lambda$ constraint obtained from S2.

\begin{table}
\centering
\small
\begin{tabular}{lcccc}
\hline
& \multicolumn{2}{c}{\textbf{S1}} & \multicolumn{2}{c}{\textbf{S14}} \\
\cline{2-5}
\textbf{Parameter} & Mean & $\sigma$ & Mean & $\sigma$ \\
\hline
$a$ (mas) & 595 & 24 & 286.3 & 3.6 \\
$e$ & 0.5560 & 0.018 & 0.9761 & 0.0037 \\
$i$ ($^\circ$) & 119.14 & 0.21 & 100.59 & 0.87 \\
$\omega$ ($^\circ$) & 122.30 & 1.4 & 334.59 & 0.87 \\
$\Omega$ ($^\circ$) & 342.04 & 0.32 & 226.38 & 0.64 \\
$t_{\rm peri}$ (yr) & 2001.80 & 0.15 & 2000.12 & 0.06 \\
\hline
\end{tabular}\hspace{1cm}
\begin{tabular}{lcc}
\hline
\textbf{Parameter} & Mean & $\sigma$ \\
\hline
$M$ ($10^6\,M_\odot$) & 4.17 & 0.19 \\
$D$ (kpc) & 8.00 & 0.19 \\
\hline
$x_0$ (mas) & $-0.2$ & $0.01$ \\
$y_0$ (mas) & $0.1$ & $0.2$ \\
$v_{x0}$ (mas yr$^{-1}$) & $0.05$ & $0.1$ \\
$v_{y0}$ (mas yr$^{-1}$) & $0.06$ & $0.1$ \\
$v_{z0}$ (km s$^{-1}$) & $0.0$ & $5.0$ \\
\hline
\end{tabular}
\caption{Gaussian priors adopted for the Keplerian orbital elements of S1 and S14 from \protect\cite{DataGillessen2016} (left). The black-hole mass and distance priors were taken from our S2 best-fit solution (Table~\ref{tab:s2params}), while the coordinate-system offsets and drift parameters were adopted from \protect\cite{Plewaoffset, DataGillessen2016} (right).}
\label{tab:otherstars}
\end{table}

Parameter inference is performed with the affine-invariant ensemble sampler implemented in \texttt{emcee} \cite{emcee}. The walkers are initialized around the Maximum A Posteriori (MAP) solution obtained from a preliminary L-BFGS-B optimization of the log-posterior \cite{optimization}. The optimizer is initialized at the prior means while enforcing the parameter bounds defined by the adopted priors.

Convergence of the chains is assessed using the integrated autocorrelation time \cite{autocorrelation}, $\tau$, for each parameter in $\Theta$. We require the chain length to satisfy $N>100\tau$ and verify the stability of the estimated autocorrelation times during the final sampling stage \cite{emcee}. We additionally monitor the mean acceptance fraction, which remains in the range $0.2$--$0.5$. After convergence, we discard a burn-in of more than $5\tau_{\rm max}$ and use the remaining samples to construct the posterior distributions and derive the constraints on $\Lambda$.

\section{Results and Discussion}\label{sec:results}

\begin{figure}
    \centering
    \includegraphics[width=\linewidth]{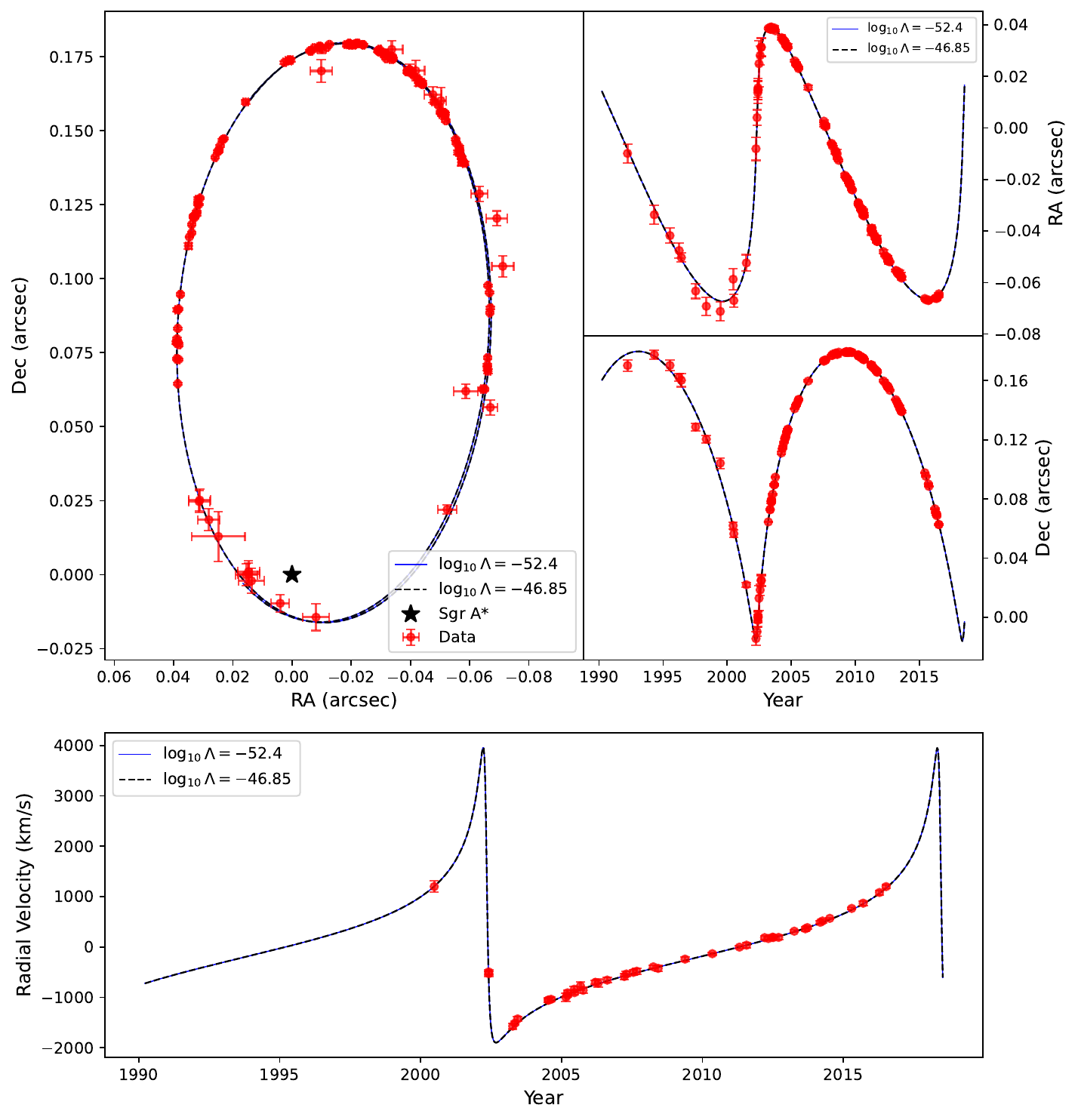}
    \caption{Astrometric and spectroscopic fits for the S2 orbit in SdS spacetime. The observational data are shown as red points with associated $1\sigma$ uncertainties. The solid curves correspond to the posterior mean model, while the dashed curves show the trajectory obtained using the 68\% upper credible bound on $\Lambda$. The left panel displays the sky-projected orbit, while the right panels show the corresponding right ascension, declination, and radial-velocity evolution.}
    \label{fig:S2orbit}
\end{figure}

\begin{figure*}
    \centering
    \includegraphics[width=\linewidth]{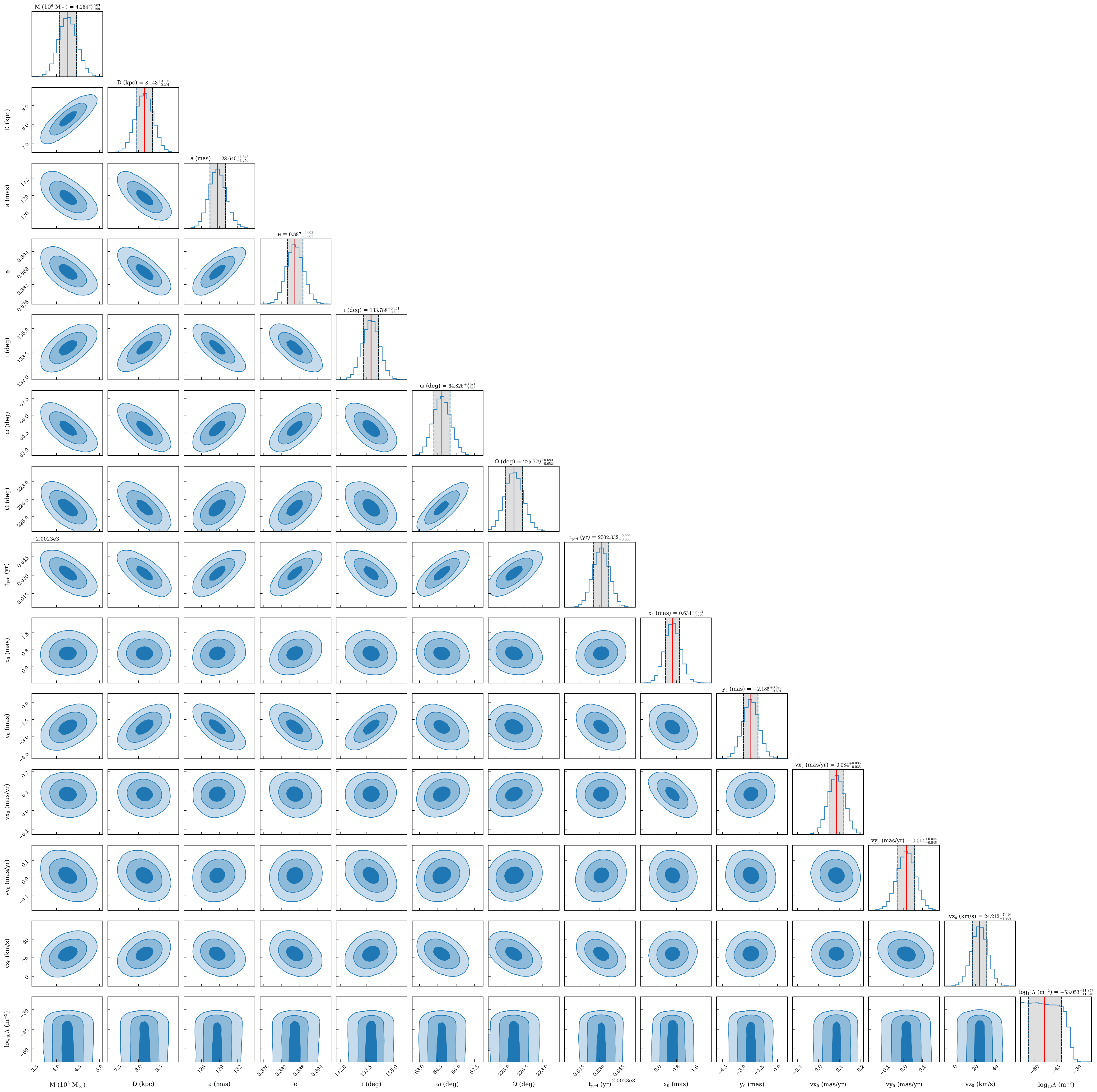}
    \caption{Marginalized posterior distribution contours with $68\%$ ($1\sigma$), $95\%$ ($2\sigma$), and $99.7\%$ ($3\sigma$) confidence levels for the parameters modeling the S2 stellar orbit in SdS spacetime. The gray-shaded regions in the diagonal histograms correspond to the $1\sigma$ credible intervals around the parameter means. The posterior mean value and its symmetric $1\sigma$ range are printed above each column, with the corresponding interval boundaries indicated in the histograms by dashed vertical lines}
    \label{fig:corner_plot_S2}
\end{figure*}

The posterior constraints obtained from the S2 analysis are summarized in Table~\ref{tab:s2params}. We recover $M=(4.27\pm0.20)\times10^6\,M_\odot$ and $D=8.14\pm0.20$~kpc, consistent with \cite{DataGillessen2016,GRAVITY2020precession}. The inferred orbital elements are likewise consistent with the literature values, showing that the SdS model provides an adequate description of the available data. The corresponding best-fit orbit and the right ascension, declination, and radial-velocity time series are shown in Figure~\ref{fig:S2orbit}, while the marginalized posterior distributions are shown in Figure~\ref{fig:corner_plot_S2}.

\begin{table}
\centering
\begin{tabular}{lcc}
\hline
\hline
Parameter & This Work & \cite{DataGillessen2016} \\
\hline

$M$ ($10^6\,M_\odot$)
& $4.267 \pm 0.201$
& $4.10 \pm 0.16$ \\

$D$ (kpc)
& $8.141 \pm 0.199$
& $8.13 \pm 0.15$ \\

$a$ (mas)
& $128.68 \pm 1.30$
& $125.5 \pm 0.9$ \\

$e$
& $0.88665 \pm 0.00259$
& $0.8839 \pm 0.0019$ \\

$i$ (deg)
& $133.78 \pm 0.45$
& $134.18 \pm 0.40$ \\

$\omega$ (deg)
& $64.84 \pm 0.66$
& $65.51 \pm 0.57$ \\

$\Omega$ (deg)
& $225.79 \pm 0.67$
& $226.94 \pm 0.60$ \\

$t_{\rm peri}$ (yr)
& $2002.332 \pm 0.006$
& $2002.33 \pm 0.01$ \\

$x_0$ (mas)
& $0.636 \pm 0.301$
& $-0.31 \pm 0.34$ \\

$y_0$ (mas)
& $-2.189 \pm 0.597$
& $-1.29 \pm 0.44$ \\

$v_{x0}$ (mas\,yr$^{-1}$)
& $0.084 \pm 0.036$
& $0.078 \pm 0.037$ \\

$v_{y0}$ (mas\,yr$^{-1}$)
& $0.013 \pm 0.045$
& $0.126 \pm 0.047$ \\

$v_{z0}$ (km\,s$^{-1}$)
& $24.17 \pm 7.12$
& $8.9 \pm 4.0$ \\

\hline

$\log_{10}(\Lambda/\mathrm{m}^{-2})$
& $-52.93 \pm 11.70$
& --- \\

\hline
\hline
\end{tabular}
\caption{
Posterior mean values and $1\sigma$ uncertainties obtained from the MCMC analysis.
The third column lists the corresponding values reported by \protect\cite{DataGillessen2016}.
}
\label{tab:s2params}
\end{table}

Similarly, analyses of S1 and S14 give orbital parameters consistent with previous determinations. In particular, the inferred values of $M$ and $D$ are consistent with those obtained from S2, despite the substantially different orbital configurations of the two stars. The resulting parameters and best-fit trajectories are given in Appendix~\ref{app:consistency}. 

The marginalized posterior distributions of $\log_{10}\Lambda$ are broad and distinctly non-Gaussian for all three stars. Interestingly, the posterior means lie near the cosmological value inferred from large-scale observations \cite{Planck2018CMB}. However, this is not statistically meaningful and should not be interpreted as a local detection of $\Lambda$. The S-stars reside in a regime where attractive gravity dominates overwhelmingly, $T_K \ll T_\Lambda$. Hence, the orbital dynamics becomes insensitive to sufficiently small values of $\Lambda$, producing a one-sided posterior with an extended low-$\Lambda$ tail. 

We therefore characterize the result through upper credible limits obtained from the cumulative marginalized posterior summarized in Table~\ref{tab:lambda_bounds}. The combined analysis gives $\Lambda \lesssim 6.9\times10^{-48} \mathrm{m}^{-2}$ at 68\% credibility and $\Lambda \lesssim 1.0\times10^{-38} \mathrm{m}^{-2}$ at 95\% credibility. Among the individual stars, S1 provides the strongest constraint, followed by S14 and S2. This is consistent with the timescale dependence \cite{Benisty2023clf} since S1's orbital period is approximately an order of magnitude longer than that of S2.
 
\begin{table*}
\centering
\begin{tabular}{lcccccc}
\hline
\hline
Star &
Median &
$68\%$ &
$\Lambda_{68\%}$ &
$95\%$ &
$\Lambda_{95\%}$ \\
\hline

S2
& $-53.18$
& $< -46.85$
& $< 1.4\times10^{-47}$
& $< -37.28$
& $< 5.2\times10^{-38}$ \\

S1
& $-53.56$
& $< -47.60$
& $< 2.5\times10^{-48}$
& $< -38.59$
& $< 2.6\times10^{-39}$ \\

S14
& $-52.92$
& $< -46.89$
& $< 1.3\times10^{-47}$
& $< -37.74$
& $< 1.8\times10^{-38}$ \\

Combined
& $-53.23$
& $< -47.16$
& $< 6.9\times10^{-48}$
& $< -37.99$
& $< 1.0\times10^{-38}$ \\

\hline
\hline
\end{tabular}
\caption{Median values and upper credible limits on $\log_{10}(\Lambda/\mathrm{m}^{-2})$ and $\Lambda$ obtained from the marginalized posterior distributions. The 68\% and 95\% bounds correspond to the cumulative posterior percentiles.}
\label{tab:lambda_bounds}
\end{table*}

\begin{figure}
    \centering
    \includegraphics[width=\linewidth]{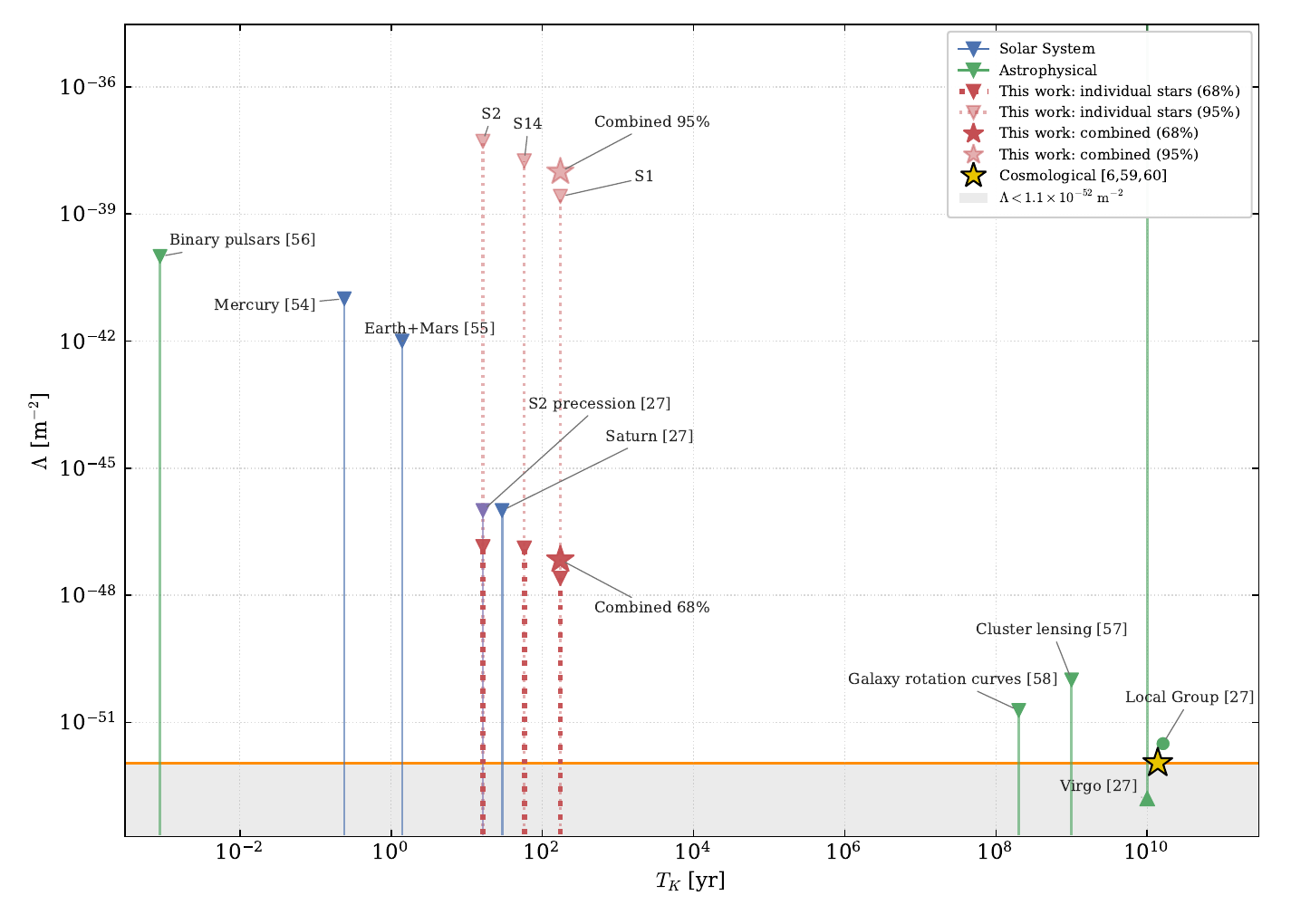}
    \caption{Constraints on the cosmological constant $\Lambda$ as a function of the characteristic dynamical timescale $T_K$ of the system. Red symbols show the new Galactic Center constraints from this work: individual stars (S2, S1, S14) are drawn as triangles with dotted drop-lines, and the combined constraint as stars; opaque (faded) markers denote 68\% (95\%) credible bounds. The yellow line marks the value $\Lambda = 1.1 \times 10^{-52}\ \mathrm{m}^{-2}$ that is consistent with the observed accelerated expansion.}
    \label{fig:Lambda_estimates}
\end{figure}

The Figure~\ref{fig:Lambda_estimates} places these bounds alongside existing constraints as a function of characteristic dynamical timescale $T_K$. The figure illustrates the expected timescale hierarchy, with constraints generally becoming tighter as $T_K$ approaches the cosmological timescale $T_\Lambda$. The bounds we obtain from S2 and S14 lie close to the constraint obtained from Saturn, consistent with their comparable dynamical timescales and, through Eq.~\eqref{kretschmann}.  

Our combined 68\% limit is the tightest constraint at comparable timescales, improving substantially over both the existing Solar-System bounds and previous GC estimates. It also improves significantly upon the precession-only S2 constraint of $\sim10^{-36}\,{\rm m}^{-2}$ obtained in section~\ref{sec:precession}. This demonstrates the additional constraining power gained by fitting the full astrometric and spectroscopic trajectory rather than the isolated periapsis-precession measurement.

The resulting sensitivity is also comparable to forecasts for pulsar timing of hypothetical S2-like pulsars around Sgr A$^*$, which reach $\Lambda\lesssim9\times10^{-47}\,{\rm m}^{-2}$ owing to the enhanced phase sensitivity of timing observations \cite{Iorio:2017auf}. We reach a similar sensitivity using existing data alone. Neural-network analyses of S2 have reported constraints at the $10^{-38}$-$10^{-40}\,{\rm m}^{-2}$ level \cite{GalikyanNN, EyamaNN}; however, these approaches rely on simplified or incomplete physical modelling of the underlying relativistic orbit. 

\section{Conclusions}\label{sec:conclusion}

In this work, we have constrained the cosmological constant using the relativistic orbits of S2, S1, and S14 around Sgr A$^*$. Our combined analysis provides the tightest upper bound on $\Lambda$ obtained at a comparable dynamical timescale, at 68\% credibility.

The stellar trajectories are modelled using \texttt{SNOPy}, our publicly available Python framework for relativistic orbit modelling and parameter inference. Using publicly available astrometric and spectroscopic observations, we performed a Bayesian parameter estimation. The stellar orbits do not provide a local detection of $\Lambda$, as the dynamics become insensitive to sufficiently small values. Instead, we obtain upper bounds on $\Lambda$ which are significantly tighter than those derived from isolated periapsis-precession estimates, demonstrating the advantage of fitting the complete relativistic orbital evolution. We also observe the expected trend that longer-period stars provide stronger sensitivity to $\Lambda$ \cite{Benisty2023clf}.

Observations of S2 by the GRAVITY collaboration provide tight constraints on the mass distribution near Sgr~A$^*$ \cite{GRAVITY2020precession}. The in-plane, prograde Schwarzschild precession of S2 has been detected at high significance, limiting plausible extended-mass distributions such as Plummer or power-law profiles within the orbit to $\sim1200\,M_\odot$ ($1\sigma$) \cite{GRAVITYextendedmass}. Stellar-population models predict masses in stars and stellar remnants to be $\sim10\,M_\odot$ near the pericenter of S2 \cite{Alexander2017,Baumgardtextendedmass,Genzel2010youthparadox}. This is comparable to the effective mass corresponding to our $95\%$ upper limit on $\Lambda$. However, it is important to note that any extended mass distribution produces retrograde precession, while $\Lambda$ produces prograde precession.

Our analysis assumes a static and spherically symmetric spacetime and neglects both black hole spin and perturbations from extended matter distributions near the GC. Longer-period S-stars and additional stellar populations in the central cluster may further enhance sensitivity to $\Lambda$ by probing longer dynamical timescales. The GC therefore occupies a unique regime different from Solar-System and cosmological scales, providing a parallel environment for placing local bounds on $\Lambda$ through precision measurements of relativistic stellar orbits around Sgr A$^*$.

\acknowledgments

The authors are grateful for discussions with Ranier Menote, David Benisty, Jenny Wagner, Ofek Birnholtz, Abdolali Banihashemi, and Davi Rodrigues. PHP acknowledges financial support from the Fundação de Amparo à Pesquisa e Inovação do Espírito Santo (FAPES, Brazil) and the Coordenação de Aperfeiçoamento de Pessoal de Nível Superior (CAPES, Brazil). WR acknowledges financial support from the CAPES. This work used the Sci-Com Lab of the Department of Physics at UFES, supported by FAPES, CAPES, and the Conselho Nacional de Desenvolvimento Científico e Tecnológico (CNPq, Brazil). DFM thanks the Research Council of Norway for their support and the resources provided by UNINETT Sigma2-the National Infrastructure for High-Performance Computing and Data Storage in Norway.


\appendix
\section{Angular shadow size and preapsis precession in SdS}\label{app:shadow}

\subsection{Angular Shadow Size}

The SdS metric possesses a black hole horizon $r_h$ and a cosmological horizon $r_c$ (with $r_h < r_c$), where $f(r_h)=f(r_c)=0$. Static observers are restricted to the region $r_h < r < r_c$ where $f(r) > 0$. Unlike asymptotically flat Schwarzschild spacetime, SdS admits no accessible asymptotic region for static observers; instead, the cosmological horizon acts as the outer causal boundary. Consequently, the large-distance behaviour of the shadow is set relative to this finite boundary rather than an observer at null infinity.

The angular radius of the shadow for a static observer at radius $r_O$ follows from the local optical geometry of null rays \cite{Perlick:2021aok} (also see \cite{sdsshadownotes} for visualizations),
\begin{equation}
    \sin^2\vartheta = \frac{(L/E)^2 g_{tt}(r_O)}{g_{\varphi\varphi}(r_O)} = \frac{b^2 f(r_O)}{r_O^2}\ .
\end{equation}
Setting $b = b_{\text{crit}}$ from \eqref{eq:bcrit} into this gives the shadow angle, 

\begin{equation}
    \sin^2\vartheta_{\text{sh}} = \frac{27(GM/c^2)^2}{r_O^2}  \frac{f(r_O)}{1 - 9\Lambda (GM/c^2)^2}.
\end{equation}

For $\Lambda = 0$, this reduces to the Schwarzschild result, where the shadow angle vanishes as $r_O \to \infty$. In SdS, however, the large-distance limit is replaced by the cosmological horizon: as $r_O \to r_c$, one has $f(r_O) \to 0$, and hence $\sin^2\vartheta_{\rm sh} \to 0$; the shadow size vanishes at the cosmological horizon. For realistic sub-horizon observers ($r_O \ll r_c$),expanding $f(r_O)$ gives, 
\begin{equation}
    \sin^2\vartheta_{\text{sh}} \approx \frac{27(GM/c^2)^2}{r_O^2(1 - 9\Lambda (GM/c^2)^2)} \left(1 - \frac{\Lambda r_O^2}{3} + \cdots\right)\ .
\end{equation}
The $\Lambda$-correction decreases the shadow size relative to the Schwarzschild case at the same coordinate radius. In the small-angle limit, the shadow radius $r_{\text{sh}} = \vartheta_{\text{sh}} r_O$ reduces to $r_{\rm sh} \approx b_{\rm crit}\sqrt{f(r_O)}$ reproducing equation \eqref{shadowrad}. 

The discussion above applies to static observers at a fixed radial coordinate $r_O$. For a comoving observer, the angular size of the shadow does not vanish at large distances as in the static case; it instead approaches a finite limiting value \cite{Perlick:2018iye, sdsshadownotes}.

\subsection{Periapsis Precession}

We give here the derivation of periapsis advance \eqref{eq:sdsprecession} following \cite{Kerrsdsprecession, Iorio:2007ub, Ruggiero:2010yn, Arakida:2012ya, Benisty2023clf}. Starting from the exact second-order geodesic equation for a timelike particle ($\epsilon = 1$), \eqref{secondordgeo}, we treat both the Schwarzschild correction ($3GMu^2/c^2$) and the $\Lambda$ term as first-order perturbations to the Kepler problem. The unperturbed (Newtonian) solution is $u_N(\varphi)=\frac{1}{p}(1+e\cos\varphi)$ with, $p \equiv \frac{L^2}{GM} = a(1-e^2)$ the semi-latus rectum of the orbit. The perturbative expansion is valid provided $\frac{GM}{c^2p}\ll1$ and $\Lambda p^2\ll1$, since both corrections already enter at first order; they can be evaluated using $u_N$ alone. Hence at first order, we replace $u^2\rightarrow u_N^2$ and $u^{-3}\rightarrow u_N^{-3}$. 

A slowly precessing orbit written as $u(\varphi)=\frac1p\left[1+e\cos\big((1-\delta)\varphi\big)\right]$ with $\delta\ll1$ the fractional shift in orbital frequency, so that the corresponding periapsis advance per orbit is $\Delta\varphi=2\pi\delta$. Substituting this ansatz into the left-hand side of the geodesic equation and expanding to first order in $\delta$ gives,
\begin{equation}
    \frac{d^2u}{d\varphi^2} + u = \frac{1}{p} + \frac{2e\delta}{p}\cos\big((1 - \delta)\varphi\big) \approx \frac{GM}{L^2} + \frac{2e\delta}{p}\cos\varphi\ .
\end{equation} 
Expressing the source terms up to linear order, 
\begin{equation}
\begin{split}
    \frac{3GM}{c^2}u_0^2\simeq\frac{3GM}{c^2p^2}+\frac{6GMe}{c^2p^2}\cos\varphi\ , \qquad 
    -\frac{\Lambda c^2}{3L^2u_0^3}\simeq-\frac{\Lambda c^2p^3}{3L^2}+\frac{\Lambda c^2p^3}{L^2}e\cos\varphi \ , 
\end{split}
\end{equation}
and collecting the resonant $\cos\varphi$ terms on both sides of the equation, we get,
\begin{equation}
    \delta=\frac{3GM}{c^2p}+\frac{\Lambda c^2p^3}{2GM}\ .
\end{equation}
The total periapsis advance per orbit is therefore,
\begin{equation}
    \Delta\varphi = 2\pi\delta = \frac{6\pi GM}{c^2p} + \frac{\pi \Lambda c^2p^3}{GM}\ ,
\end{equation}
which matches \eqref{eq:sdsprecession}.

\section{Orbit Modelling using SNOPy}\label{app:orbitmodel}

Here we present the relativistic orbit modelling framework implemented in \texttt{SNOPy}\footnote{\url{https://github.com/PrajwalPrem/SNOPy}}. Stellar motion is treated as a timelike test-particle geodesic in a general spherically symmetric spacetime, applied here to the SdS metric. Geodesic trajectories are then mapped to the observer frame to evaluate the astrometric and spectroscopic observables used in our likelihood analysis\footnote{See also \texttt{GC-OrbitFit}\footnote{\url{https://github.com/pmplewa/GC-OrbitFit}}, which uses nested sampling for inference and the \texttt{PyGRO} package \cite{Pygropaper} for general numerical integration of the massive and massless geodesic equations.}. 

\begin{figure}
    \centering
    \includegraphics[width=0.7\linewidth]{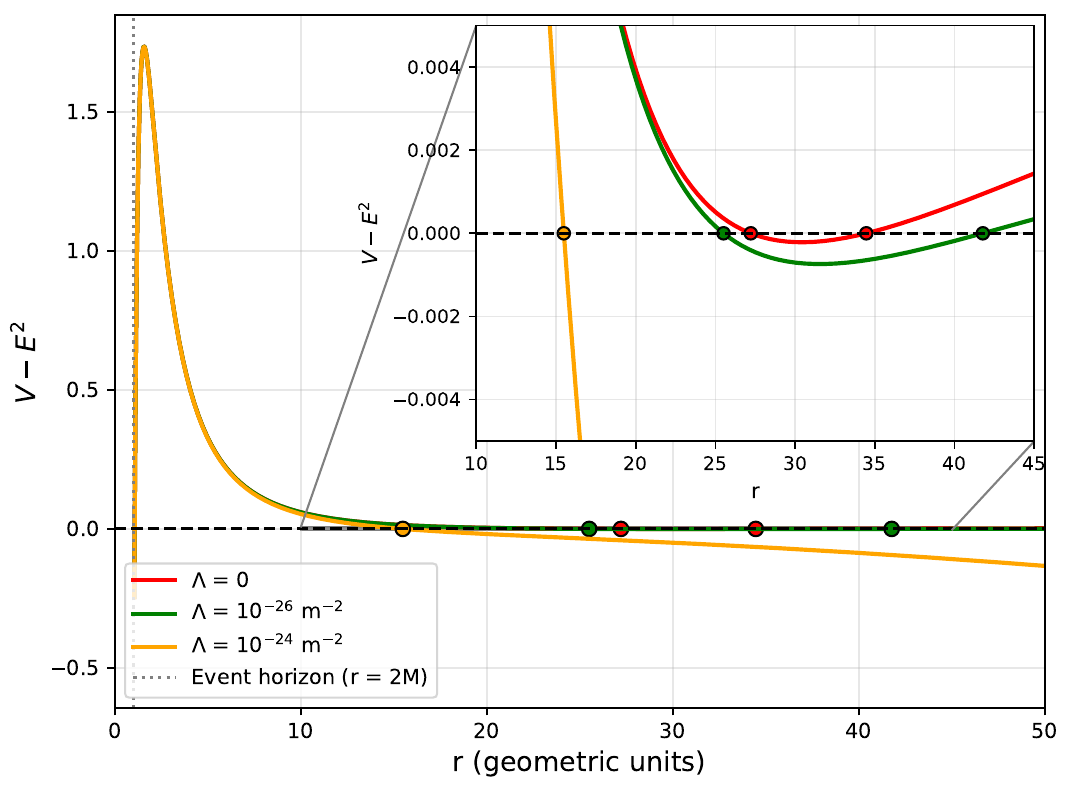}
    \caption{Effective potential $V(r)$ in SdS spacetime for different values of $\Lambda$. Bound motion occurs between the turning points $r_p$ and $r_a$, which are highlighted with $\bullet$. For sufficiently large $\Lambda$ (yellow curve), the outer potential barrier disappears, leading to unbound trajectories.}
    \label{fig:sdsturningpoints}
\end{figure}

The radial motion is governed by the effective potential $V(r)$ appearing in \eqref{radialnormalization}. Bound motion exists when the particle is confined between two turning points, corresponding to the pericenter $r_p$ and the apocenter $r_a$, where $\dot{r} = 0$. The structure of the effective potential for different values of $\Lambda$ is illustrated in Figure~\ref{fig:sdsturningpoints}. For sufficiently large $\Lambda$, the outer potential barrier is suppressed and bound orbits cease to exist. 

The trajectory is obtained by integrating the geodesic equations \eqref{geodesic} where the Christoffel symbols $\Gamma^\mu_{\alpha\beta}$ are computed from the SdS metric \eqref{sdsmetric}. We initialize the orbit at apocenter at the proper time $\tau = \tau_a$ in the equatorial plane ($\theta = \pi/2$). The initial phase-space vector is therefore,
\begin{equation}
    \{x^\mu, \ \dot{x}^\mu \}_{\tau_a} = \{t_a, \ r_a, \ \pi/2, \ \varphi_a, \ \dot{t}_a, \ 0, \ 0, \ \dot{\varphi}_a\}.
\end{equation}
The conserved $E$ and $L$ are uniquely determined by the semi-major axis $a$ and eccentricity $e$. By evaluating the normalization condition \eqref{radialnormalization} at the apsides $r_a = a(1+e)$ and $r_p = a(1-e)$, we find,
\begin{equation}
    E = \sqrt{\frac{f(r_a) f(r_p) (r_a^2 - r_p^2)}{r_a^2 f(r_p) - r_p^2 f(r_a)}}, \quad L = \sqrt{\frac{r_a^2 r_p^2 (f(r_p) - f(r_a))}{r_a^2 f(r_p) - r_p^2 f(r_a)}}.
\end{equation}
The initial velocities then follow from the conserved quantities through $\dot{t}_a = E/f(r_a)$ and $\dot{\varphi}_a = L/r_a^2$. The orbit is integrated both forward and backwards in proper time from $\tau_a$ using the adaptive high-order Runge-Kutta solver DOP853 \cite{hairer1993ode1}.

\subsection{Projection to Observer’s Sky}

The numerical integration of the geodesic equations yields the stellar trajectory $\{t(\tau), r(\tau), \varphi(\tau)\}$ along with the velocity components $u^\mu = \{u^t(\tau), u^r(\tau), u^\varphi(\tau)\}$. In the orbital plane $(z_{\rm orb} = 0)$, the Cartesian orbital position $\{x_{\rm orb}, y_{\rm orb}\}$ and velocity $\{v_{\rm orb}^x, v_{\rm orb}^y\}$ components are,
\begin{equation}
\begin{split}
    x_{\rm orb} = r\cos\varphi, &\qquad y_{\rm orb} = r\sin\varphi\ ,\\
    v_{\rm orb}^x = v_r \cos \varphi - v_\varphi \sin \varphi\ ,&\qquad  v_{\rm orb}^y = v_r \sin \varphi + v_\varphi \cos \varphi \ ,
\end{split}
\end{equation}
where $v_r = u^r/u^t$ and $v_\varphi = r u^\varphi/u^t$ are the radial and azimuthal coordinate velocities, respectively. The transformation from the orbital plane to the observer sky frame is performed using the standard Thiele–Innes parametrization, specified by the inclination $i$, longitude of the ascending node $\Omega$, and argument of periapsis $\omega$. The corresponding Thiele–Innes elements (also called Innes constants) are \cite{doublestarsbook},
\begin{equation}
\begin{split}
    A &= \cos\Omega\cos\omega - \sin\Omega\sin\omega\cos i, \\
    B &= \sin\Omega\cos\omega + \cos\Omega\sin\omega\cos i, \\
    C &= -\sin\omega\sin i, \\
    F &= -\cos\Omega\sin\omega - \sin\Omega\cos\omega\cos i, \\
    G &= -\sin\Omega\sin\omega + \cos\Omega\cos\omega\cos i, \\
    H &= -\cos\omega\sin i\ .
\end{split}
\end{equation}
The projected sky coordinates and velocities are then obtained through
\begin{equation}
    \begin{bmatrix}
    x^{\rm sky} \\
    y^{\rm sky} \\
    z^{\rm sky}
    \end{bmatrix}
    =
    \mathcal{R}
    \begin{bmatrix}
    x_{\rm orb} \\
    y_{\rm orb}
    \end{bmatrix}, \qquad 
    \begin{bmatrix}
    v_x^{\rm sky} \\
    v_y^{\rm sky} \\
    -v_z^{\rm sky}
    \end{bmatrix}
    =
    \mathcal{R}
    \begin{bmatrix}
    v_{\rm orb}^x \\
    v_{\rm orb}^y
    \end{bmatrix},
\end{equation}
where the matrix $\mathcal{R}$ is given by,
\begin{equation}
    \mathcal{R}=
    \begin{bmatrix}
        B & G \\
        A & F \\
        C & H
    \end{bmatrix}.
\end{equation}
Here, $\{x^{\rm sky}, y^{\rm sky}\}$ denote the projected coordinates on the plane of the sky, while $z^{\rm sky}$ is the line-of-sight displacement relative to Sgr A$^*$. We adopt the convention where $z^{\rm sky}$ points toward the observer, such that $v_z^{\rm sky} > 0$ corresponds to approaching (blueshifted) motion. Finally, the observable angular coordinates, namely the Right Ascension $(\alpha)$ and Declination $(\delta)$, are obtained through the geometric projection, 
\begin{equation}
    \alpha = \tan^{-1} \left( \frac{x^{\rm sky}}{D - z^{\rm sky}} \right), \quad \delta = \tan^{-1} \left( \frac{y^{\rm sky}}{D - z^{\rm sky}} \right),
\end{equation}
where $D$ is the distance to Sgr A$^*$. While the flat-sky approximation ($\alpha \simeq x^{\rm sky}/D$) is typically sufficient, we retain the transcendental form.

\subsection{R{\o}mer delay}

The varying line-of-sight position of the S-stars along their orbits induces a modulation in the light-travel time to the observer. This geometric effect gives rise to the R{\o}mer delay, relating the emission time $t_{\rm em}$ to the observation time $t_{\rm obs}$ recorded at Earth through
\begin{equation}\label{romerdelay}
    t_{\rm em} = t_{\rm obs} - \frac{z^{\rm sky}(t_{\rm em})}{c}\ .
\end{equation}
Since the stellar position $z^{\rm sky}$ depends on the orbital phase at emission, the equation must be solved iteratively at each observational epoch to determine the corresponding emission time \cite{Shaymatov2023jfa}. 

We neglect the Shapiro delay in the present analysis. For the S-stars considered here, its contribution remains at most of order a few minutes, which is well below the temporal resolution of the available observations.

\subsection{Reference Frame Motion}

The infrared astrometric frame is tied to the radio reference frame of Sgr A$^*$ via SiO maser emission from nearby red giant stars, leading to small residual offsets and drifts relative to the black-hole rest frame \cite{Plewaoffset}. To account for these effects, we include nuisance parameters corresponding to constant positional offsets $(x_0,y_0)$ and linear proper-motion drifts $(v_{x0},v_{y0})$ of the central mass. The theoretical sky coordinates evaluated at the observation time $t_{\rm obs}$ are therefore written as \cite{Keckredshift}, 
\begin{equation}
\begin{split}
    \alpha^{\rm th}(t_{\rm obs}) &= \alpha(t_{\rm em}) + x_0 + v_{x0}(t_{\rm em} - t_{\rm ref}), \\
    \delta^{\rm th}(t_{\rm obs}) &= \delta(t_{\rm em}) + y_0 + v_{y0}(t_{\rm em} - t_{\rm ref}),
\end{split}
\end{equation}
where, $t_{\rm ref}$ denotes a reference epoch. In this work, it is chosen to coincide with the epoch of pericenter passage $t_{p}$ for each star. 

Similarly, the spectroscopic frame may contain a constant line-of-sight velocity offset. This is modelled through the nuisance parameter $v_{z0}$, which absorbs residual uncertainties associated with the Local Standard of Rest (LSR) correction and spectroscopic corrections.

\subsection{Relativistic Redshift}

The observed spectroscopic redshift contains contributions from both the relativistic Doppler effect and gravitational time dilation. For a star moving in the SdS spacetime, the total redshift measured by a static distant observer is \cite{Grould2017bsw}, 
\begin{equation}
    1 + \zeta_{\rm tot} = \frac{1}{\sqrt{f(r_{\rm em})}} \frac{\sqrt{1 - v^2/c^2}}{1 - v_{z}^{\rm sky}/c}\ ,
\end{equation}
where $f(r_{\rm em})$ is the metric function \eqref{sdsmetric} evaluated at the emission radius, $v$ is the total three-velocity magnitude, and $v_{z}^{\rm sky}$ is the line-of-sight velocity component in the sky-projected frame. The first factor represents the gravitational redshift in SdS spacetime, while the second incorporates the special-relativistic Doppler factor. The radial velocity is then given by, $v_{R}^{\rm th} = c  \zeta_{\rm tot} + v_{z0}$, where $v_{z0}$ denotes the constant velocity offset defined earlier.

To summarize, the numerical integration of the geodesic equations, together with the sky-plane projection, R{\o}mer delay correction, reference-frame offsets, and relativistic redshift correction, provides the complete set of theoretical observables as a function of model parameters. The model therefore predicts the astrometric coordinates in right ascension and declination $(\alpha,\delta) \equiv ({\rm RA}^{\rm th}, {\rm Dec}^{\rm th})$ and the radial velocity $v_R^{\rm th}$. These quantities are subsequently assembled into the theoretical data vector used in the statistical analysis.


\section{Results for S1 and S14}\label{app:consistency}

Here we present the full posterior constraints obtained from the analyses of the S1 and S14 trajectories. These stars probe very different orbital regimes: S14 follows a highly eccentric orbit, whereas S1 traces a longer-period orbit. The currently available data for these stars are insufficient to independently constrain the global parameters. Therefore, we adopt Gaussian priors on $M$ and $D$, informed by the S2 analysis (see Table~\ref{tab:otherstars}), and use the trajectories of these stars to obtain additional constraints on $\Lambda$.

\begin{figure}
    \centering
    \includegraphics[width=0.9\linewidth]{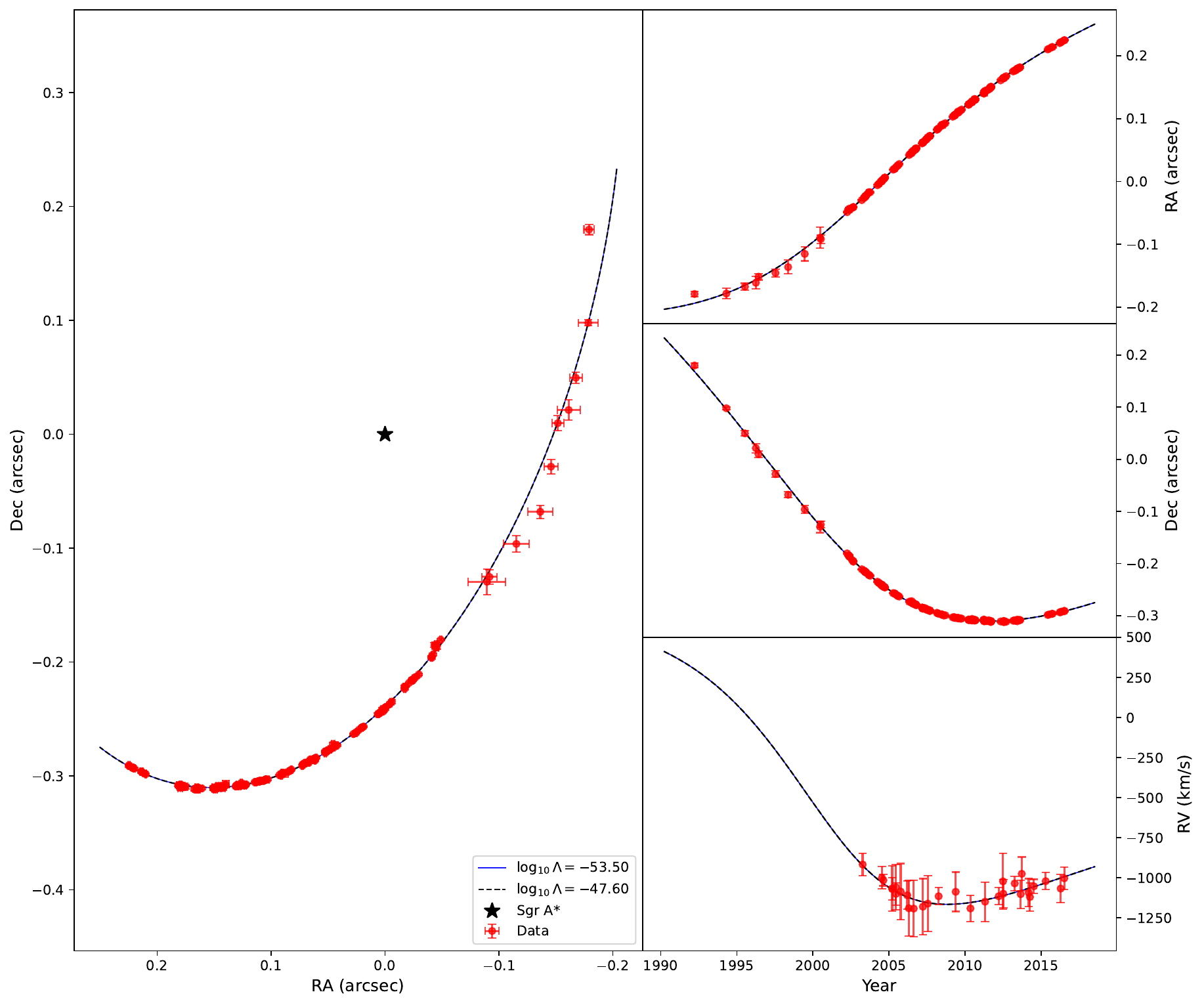}
    \caption{Best-fit SdS orbit for S1. The figure shows the projected sky trajectory together (left) with the corresponding right ascension, declination, and radial-velocity fits (right) to the observational data. As in Figure~\protect\ref{fig:S2orbit}, both the posterior mean trajectory and the orbit corresponding to the 68\% upper credible bound on $\Lambda$ are shown.}
    \label{fig:S1orbit}
\end{figure}

\begin{figure}
    \centering
    \includegraphics[width=0.9\linewidth]{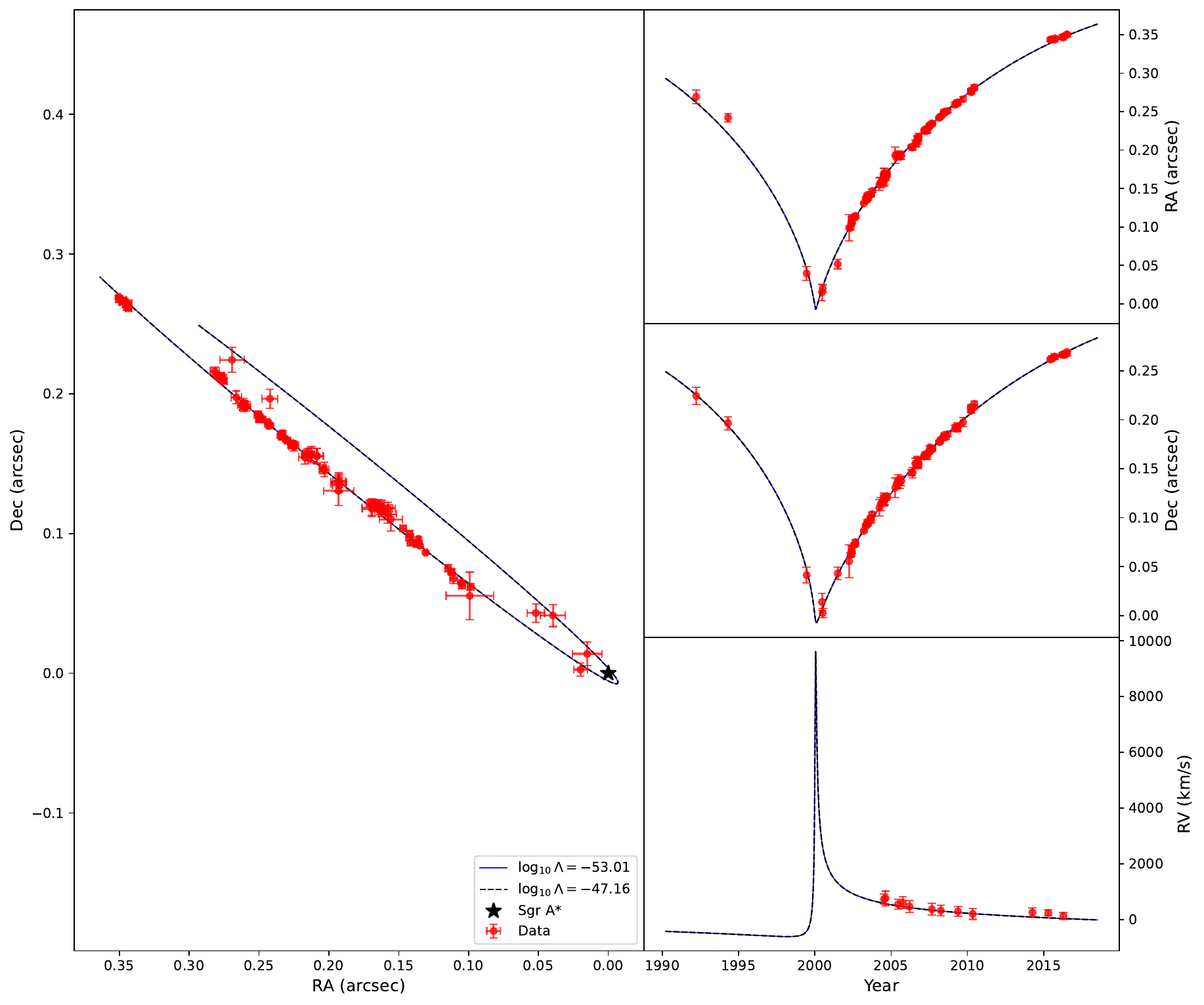}
    \caption{Same as Figure~\ref{fig:S1orbit}, but for the S14 star.}
    \label{fig:S14orbit}
\end{figure}

\begin{figure}
    \centering
    \includegraphics[width=0.8\linewidth]{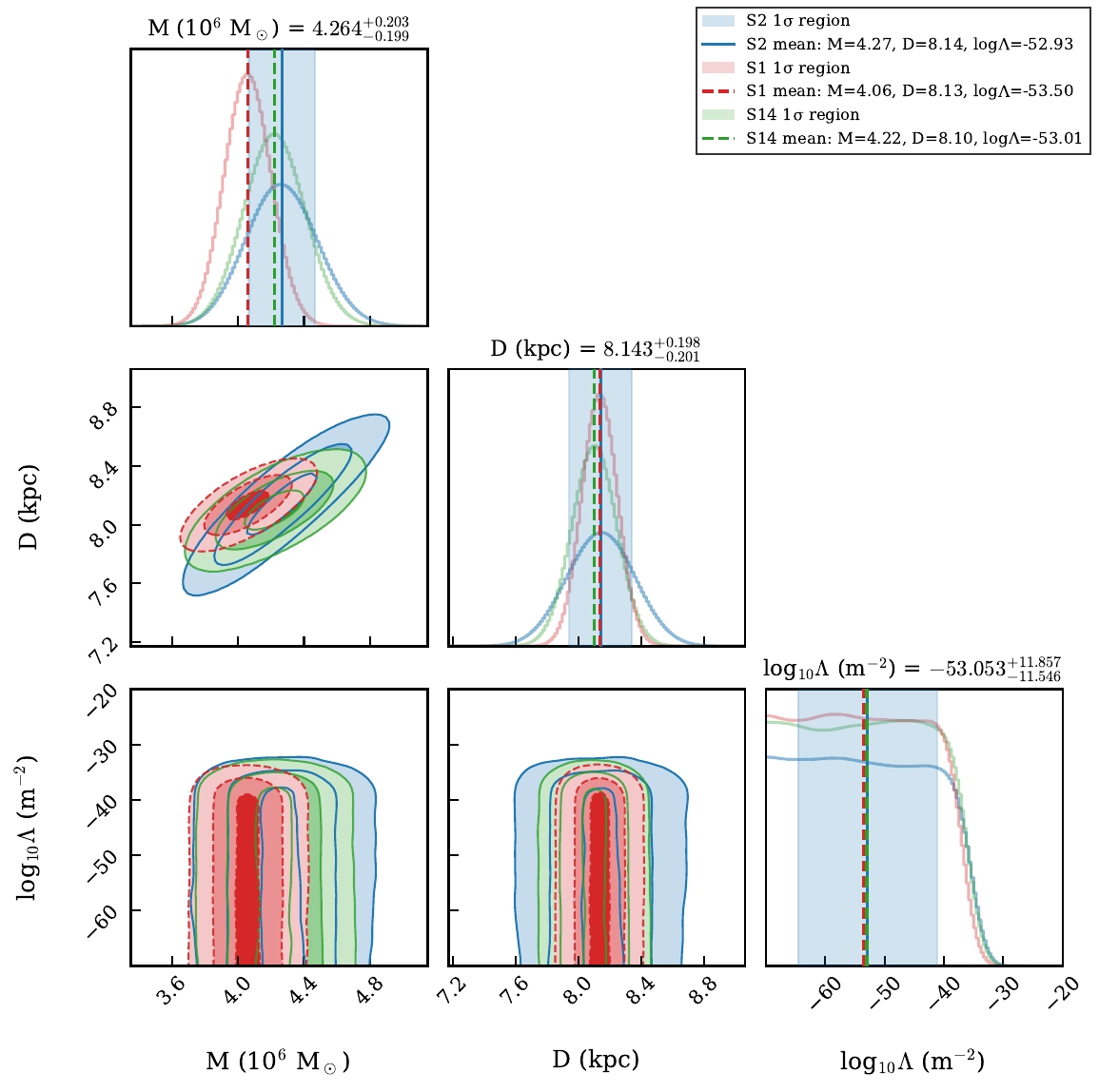}
    \caption{Marginalized posterior distributions for $M, D$ and $\Lambda$ obtained from the S2, S1, and S14 orbital analyses (in blue, red and green respectively). The contours indicate the $1\sigma$, $2\sigma$ and $3\sigma$ confidence levels for the parameters.}
    \label{fig:corner_consistency}
\end{figure}

\begin{figure}
    \centering
    \includegraphics[width=\linewidth]{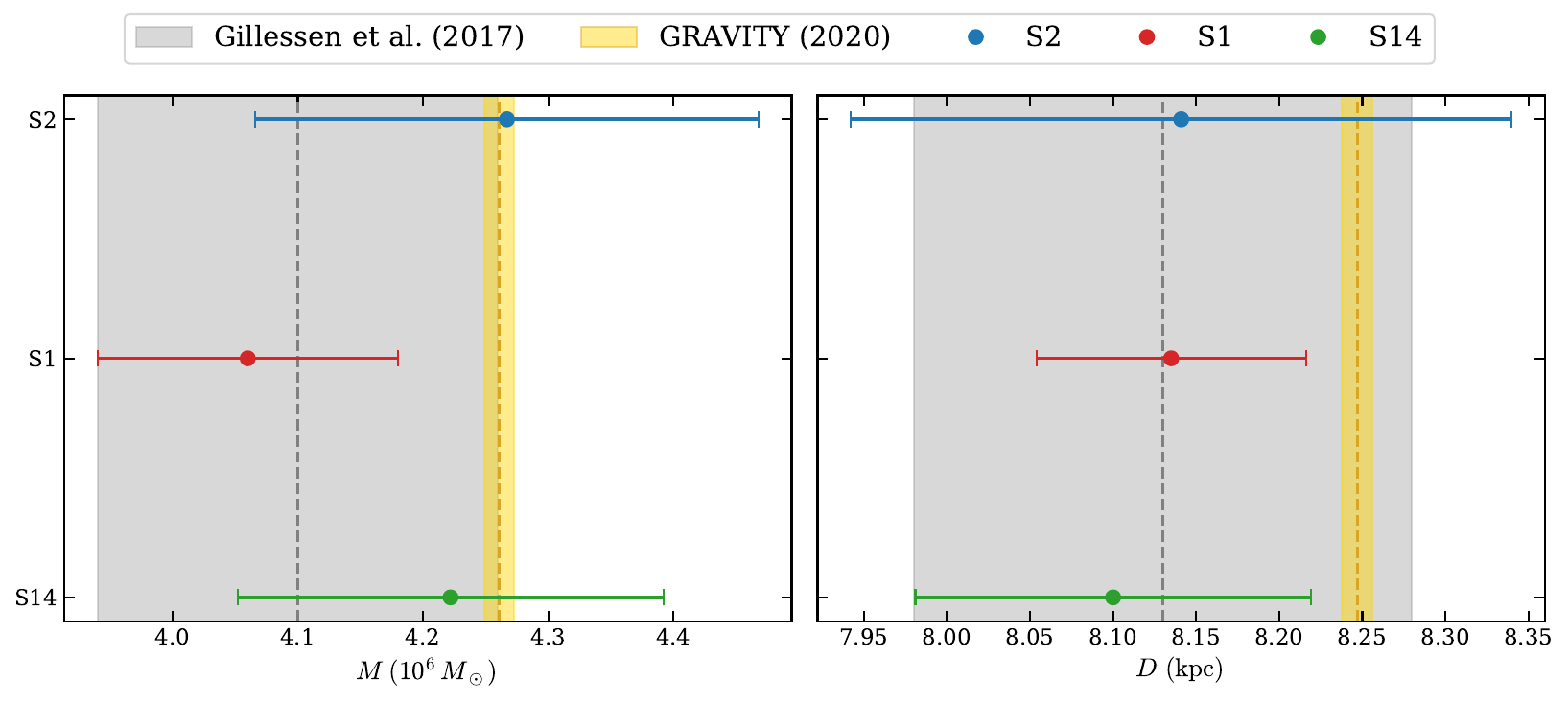}
    \caption{Comparison of the inferred black-hole mass $M$ and Galactic Center distance $D$ obtained in this work from S2, S1, and S14 with previous measurements from \protect\cite{DataGillessen2016} and \protect\cite{GRAVITY2020precession}. Error bars correspond to $1\sigma$ uncertainties.}
    \label{fig:Mass_Distance_comparison}
\end{figure}

The complete posterior parameter estimates for S1 and S14 are listed in Table~\ref{tab:s1_s14_parameters}. The corresponding best-fit orbital trajectories and radial-velocity fits are shown in Figs.~\ref{fig:S1orbit} and \ref{fig:S14orbit}. The inferred values of the parameters are consistent, within uncertainties, with the estimates of \cite{DataGillessen2016}. In particular, $M$ and $D$ from S1 and S14 agree with those derived from the S2 analysis. The corresponding posterior distributions are shown in Figure~\ref{fig:corner_consistency}, while Figure~\ref{fig:Mass_Distance_comparison} compares our inferred values with the orbital analyses of \cite{DataGillessen2016} and \cite{GRAVITY2020precession}. The reduced chi-squared values are $\chi_\nu^2 = 0.91$ for S2, $\chi_\nu^2 = 0.75$ for S1 and $\chi_\nu^2 = 1.03$ for S14, indicating that the relativistic SdS framework provides an adequate description of the astrometric and spectroscopic observations within the reported uncertainties. 


\begin{table}
\centering
\begin{tabular}{lcc}
\hline
\hline
Parameter & S1 & S14 \\
\hline

$M$ ($10^6\,M_\odot$)
& $4.060 \pm 0.120$
& $4.222 \pm 0.170$ \\

$D$ (kpc)
& $8.135 \pm 0.081$
& $8.100 \pm 0.119$ \\

$a$ (mas)
& $609.63 \pm 11.01$
& $298.17 \pm 2.57$ \\

$e$
& $0.57054 \pm 0.00781$
& $0.96887 \pm 0.00151$ \\

$i$ (deg)
& $119.685 \pm 0.137$
& $99.985 \pm 0.403$ \\

$\omega$ (deg)
& $121.808 \pm 0.596$
& $335.247 \pm 0.690$ \\

$\Omega$ (deg)
& $342.409 \pm 0.197$
& $226.831 \pm 0.319$ \\

$t_{\rm peri}$ (yr)
& $2001.648 \pm 0.072$
& $2000.037 \pm 0.030$ \\

$x_0$ (mas)
& $-0.200 \pm 0.010$
& $-0.200 \pm 0.010$ \\

$y_0$ (mas)
& $0.089 \pm 0.199$
& $0.107 \pm 0.196$ \\

$v_{x0}$ (mas\,yr$^{-1}$)
& $0.024 \pm 0.076$
& $0.116 \pm 0.084$ \\

$v_{y0}$ (mas\,yr$^{-1}$)
& $0.103 \pm 0.094$
& $0.084 \pm 0.086$ \\

$v_{z0}$ (km\,s$^{-1}$)
& $1.49 \pm 4.92$
& $1.10 \pm 4.98$ \\

$\log_{10}(\Lambda/\mathrm{m}^{-2})$
& $-53.50 \pm 11.23$
& $-53.01 \pm 11.55$ \\

\hline
\hline
\end{tabular}

\caption{Posterior mean values and corresponding $1\sigma$ uncertainties for S1 and S14}

\label{tab:s1_s14_parameters}
\end{table}

\paragraph{Data Availability} The data utilized in this study are publicly available from \cite{DataGillessen2016} and are accessible electronically through the catalogue \cite{catalogue}. We performed the orbit modelling and Bayesian analysis using our publicly available Python framework \texttt{SNOPy}\footnote{\url{https://github.com/PrajwalPrem/SNOPy}}.

\bibliographystyle{JHEP}
\bibliography{example}

@article{GRAVITY2018redshift,
    author = "Abuter, R. and others",
    collaboration = "GRAVITY",
    title = "{Detection of the gravitational redshift in the orbit of the star S2 near the Galactic centre massive black hole}",
    eprint = "1807.09409",
    archivePrefix = "arXiv",
    primaryClass = "astro-ph.GA",
    doi = "10.1051/0004-6361/201833718",
    journal = "Astron. Astrophys.",
    volume = "615",
    pages = "L15",
    year = "2018"
}

@article{GRAVITY2020precession,
    author = "Abuter, R. and others",
    collaboration = "GRAVITY",
    title = "{Detection of the Schwarzschild precession in the orbit of the star S2 near the Galactic centre massive black hole}",
    eprint = "2004.07187",
    archivePrefix = "arXiv",
    primaryClass = "astro-ph.GA",
    doi = "10.1051/0004-6361/202037813",
    journal = "Astron. Astrophys.",
    volume = "636",
    pages = "L5",
    year = "2020"
}

@article{Alexander2017,
   title={Stellar Dynamics and Stellar Phenomena Near a Massive Black Hole},
   volume={55},
   ISSN={1545-4282},
   url={http://dx.doi.org/10.1146/annurev-astro-091916-055306},
   DOI={10.1146/annurev-astro-091916-055306},
   number={1},
   journal={Annual Review of Astronomy and Astrophysics},
   publisher={Annual Reviews},
   author={Alexander, Tal},
   year={2017},
   month=Aug, pages={17–57} }

@article{GRAVITYextendedmass,
    author = "Abd El Dayem, Karim and others",
    collaboration = "GRAVITY",
    title = "{Improving constraints on the extended mass distribution in the Galactic center with stellar orbits}",
    eprint = "2409.12261",
    archivePrefix = "arXiv",
    primaryClass = "astro-ph.GA",
    doi = "10.1051/0004-6361/202452274",
    journal = "Astron. Astrophys.",
    volume = "692",
    pages = "A242",
    year = "2024"
}

@article{Kagramanova:2006ax,
    author = "Kagramanova, Valeria and Kunz, Jutta and Lammerzahl, Claus",
    title = "{Solar system effects in Schwarzschild-de Sitter spacetime}",
    eprint = "gr-qc/0602002",
    archivePrefix = "arXiv",
    doi = "10.1016/j.physletb.2006.01.069",
    journal = "Phys. Lett. B",
    volume = "634",
    pages = "465--470",
    year = "2006"
}

@article{Ishak:2007ea,
    author = "Ishak, Mustapha and Rindler, Wolfgang and Dossett, Jason and Moldenhauer, Jacob and Allison, Chris",
    title = "{A New Independent Limit on the Cosmological Constant/Dark Energy from the Relativistic Bending of Light by Galaxies and Clusters of Galaxies}",
    eprint = "0710.4726",
    archivePrefix = "arXiv",
    primaryClass = "astro-ph",
    doi = "10.1111/j.1365-2966.2008.13468.x",
    journal = "Mon. Not. Roy. Astron. Soc.",
    volume = "388",
    pages = "1279--1283",
    year = "2008"
}

@article{Iorio:2017auf,
    author = "Iorio, Lorenzo",
    title = "{Perspectives on constraining a cosmological constant-type parameter with pulsar timing in the Galactic Center}",
    eprint = "1712.09106",
    archivePrefix = "arXiv",
    primaryClass = "gr-qc",
    doi = "10.3390/universe4040059",
    journal = "Universe",
    volume = "4",
    number = "4",
    pages = "59",
    year = "2018"
}

@article{Sereno:2006re,
    author = "Sereno, Mauro and Jetzer, Philippe",
    title = "{Solar and stellar system tests of the cosmological constant}",
    eprint = "astro-ph/0602438",
    archivePrefix = "arXiv",
    doi = "10.1103/PhysRevD.73.063004",
    journal = "Phys. Rev. D",
    volume = "73",
    pages = "063004",
    year = "2006"
}

@article{Jetzer:2006gn,
    author = "Jetzer, Philippe and Sereno, Mauro",
    title = "{Two-body problem with the cosmological constant and observational constraints}",
    eprint = "astro-ph/0601612",
    archivePrefix = "arXiv",
    doi = "10.1103/PhysRevD.73.044015",
    journal = "Phys. Rev. D",
    volume = "73",
    pages = "044015",
    year = "2006"
}

@article{Vagnozzi2022horizon,
    author = "Vagnozzi, Sunny and others",
    title = "{Horizon-scale tests of gravity theories and fundamental physics from the Event Horizon Telescope image of Sagittarius A}",
    eprint = "2205.07787",
    archivePrefix = "arXiv",
    primaryClass = "gr-qc",
    reportNumber = "UCI-HEP-TR-2022-07",
    doi = "10.1088/1361-6382/acd97b",
    journal = "Class. Quant. Grav.",
    volume = "40",
    number = "16",
    pages = "165007",
    year = "2023"
}

@article{EventHorizonTelescope2022wkp,
    author = "Akiyama, Kazunori and others",
    collaboration = "Event Horizon Telescope",
    title = "{First Sagittarius A* Event Horizon Telescope Results. I. The Shadow of the Supermassive Black Hole in the Center of the Milky Way}",
    eprint = "2311.08680",
    archivePrefix = "arXiv",
    primaryClass = "astro-ph.HE",
    doi = "10.3847/2041-8213/ac6674",
    journal = "Astrophys. J. Lett.",
    volume = "930",
    number = "2",
    pages = "L12",
    year = "2022"
}

@article{EventHorizonTelescope2019dse,
    author = "Akiyama, Kazunori and others",
    collaboration = "Event Horizon Telescope",
    title = "{First M87 Event Horizon Telescope Results. I. The Shadow of the Supermassive Black Hole}",
    eprint = "1906.11238",
    archivePrefix = "arXiv",
    primaryClass = "astro-ph.GA",
    doi = "10.3847/2041-8213/ab0ec7",
    journal = "Astrophys. J. Lett.",
    volume = "875",
    pages = "L1",
    year = "2019"
}

@article{KeckGhez,
    author = "Ghez, A. M. and Salim, S. and Hornstein, Seth D. and Tanner, A. and Morris, M. and Becklin, E. E. and Duchene, G.",
    title = "{Stellar orbits around the galactic center black hole}",
    eprint = "astro-ph/0306130",
    archivePrefix = "arXiv",
    doi = "10.1086/427175",
    journal = "Astrophys. J.",
    volume = "620",
    pages = "744--757",
    year = "2005"
}

@article{DataGillessen2016,
    author = "Gillessen, S. and others",
    title = "{An Update on Monitoring Stellar Orbits in the Galactic Center}",
    doi = "10.3847/1538-4357/aa5c41",
    journal = "Astrophys. J.",
    volume = "837",
    number = "1",
    pages = "30",
    year = "2017"
}

@article{GRAVITY2021Massdist,
    author = "Abuter, R. and others",
    collaboration = "GRAVITY",
    title = "{Mass distribution in the Galactic Center based on interferometric astrometry of multiple stellar orbits}",
    eprint = "2112.07478",
    archivePrefix = "arXiv",
    primaryClass = "astro-ph.GA",
    doi = "10.1051/0004-6361/202142465",
    journal = "Astron. Astrophys.",
    volume = "657",
    pages = "L12",
    year = "2022"
}

@article{KeckHees2017,
    author = "Hees, A. and others",
    title = "{Testing General Relativity with stellar orbits around the supermassive black hole in our Galactic center}",
    eprint = "1705.07902",
    archivePrefix = "arXiv",
    primaryClass = "astro-ph.GA",
    doi = "10.1103/PhysRevLett.118.211101",
    journal = "Phys. Rev. Lett.",
    volume = "118",
    number = "21",
    pages = "211101",
    year = "2017"
}

@article{Baumgardtextendedmass,
   title={The distribution of stars around the Milky Way’s central black hole: III. Comparison with simulations},
   volume={609},
   ISSN={1432-0746},
   url={http://dx.doi.org/10.1051/0004-6361/201730462},
   DOI={10.1051/0004-6361/201730462},
   journal={Astronomy \&; Astrophysics},
   publisher={EDP Sciences},
   author={Baumgardt, H. and Amaro-Seoane, P. and Schödel, R.},
   year={2017},
   month=Dec, pages={A28} }

@article{Genzel2010youthparadox,
  title={The galactic center massive black hole and nuclear star cluster},
  author={Reinhard Genzel and Frank Eisenhauer and Stefan Gillessen},
  journal={Reviews of Modern Physics},
  year={2010},
  volume={82},
  pages={3121-3195},
  url={https://api.semanticscholar.org/CorpusID:119195974}
}

@article{fRS2,
    author = "de Martino, Ivan and della Monica, Riccardo and de Laurentis, Mariafelicia",
    title = "{f(R) gravity after the detection of the orbital precession of the S2 star around the Galactic Center massive black hole}",
    eprint = "2106.06821",
    archivePrefix = "arXiv",
    primaryClass = "gr-qc",
    doi = "10.1103/PhysRevD.104.L101502",
    journal = "Phys. Rev. D",
    volume = "104",
    number = "10",
    pages = "L101502",
    year = "2021"
}

@article{scalartensorvector,
    author = "Della Monica, Riccardo and de Martino, Ivan and de Laurentis, Mariafelicia",
    title = "{Orbital precession of the S2 star in Scalar{\textendash}Tensor{\textendash}Vector Gravity}",
    eprint = "2105.12687",
    archivePrefix = "arXiv",
    primaryClass = "gr-qc",
    doi = "10.1093/mnras/stab3727",
    journal = "Mon. Not. Roy. Astron. Soc.",
    volume = "510",
    number = "4",
    pages = "4757--4766",
    year = "2022"
}

@article{scalartensorvectorMOG,
    author = "Monica, Riccardo Della and de Martino, Ivan and de Laurentis, Mariafelicia",
    title = "{Constraining MOdified Gravity with the S2 Star}",
    eprint = "2206.12699",
    archivePrefix = "arXiv",
    primaryClass = "gr-qc",
    doi = "10.3390/universe8020137",
    journal = "Universe",
    volume = "8",
    number = "2",
    pages = "137",
    year = "2022"
}

@article{HorndeskiS2,
    author = "Della Monica, Riccardo and de Martino, Ivan and Vernieri, Daniele and de Laurentis, Mariafelicia",
    title = "{Testing horndeski gravity with S2 star orbit}",
    eprint = "2212.05082",
    archivePrefix = "arXiv",
    primaryClass = "gr-qc",
    doi = "10.1093/mnras/stac3648",
    journal = "Mon. Not. Roy. Astron. Soc.",
    volume = "519",
    number = "2",
    pages = "1981--1988",
    year = "2022"
}

@article{beyondHorndeski,
    author = "Zhang, Zelin and Chen, Songbai and Jing, Jiliang",
    title = "{Constraining a disformal Schwarzschild black hole in DHOST theories with the orbit of the S2 star}",
    eprint = "2404.05957",
    archivePrefix = "arXiv",
    primaryClass = "gr-qc",
    doi = "10.1140/epjc/s10052-024-13179-6",
    journal = "Eur. Phys. J. C",
    volume = "84",
    number = "8",
    pages = "827",
    year = "2024"
}

@article{unveilingwormhole,
    author = "Della Monica, Riccardo and de Martino, Ivan",
    title = "{Unveiling the nature of SgrA* with the geodesic motion of S-stars}",
    eprint = "2112.01888",
    archivePrefix = "arXiv",
    primaryClass = "astro-ph.GA",
    doi = "10.1088/1475-7516/2022/03/007",
    journal = "JCAP",
    volume = "03",
    number = "03",
    pages = "007",
    year = "2022"
}

@article{nakedsingularity,
    author = "Bambhaniya, Parth and Joshi, Ashok B. and Dey, Dipanjan and Joshi, Pankaj S. and Mazumdar, Arindam and Harada, Tomohiro and Nakao, Ken-ichi",
    title = "{Relativistic orbits of S2 star in the presence of scalar field}",
    eprint = "2209.12610",
    archivePrefix = "arXiv",
    primaryClass = "gr-qc",
    reportNumber = "RUP-22-20, AP-GR-183, NITEP 146",
    doi = "10.1140/epjc/s10052-024-12477-3",
    journal = "Eur. Phys. J. C",
    volume = "84",
    number = "2",
    pages = "124",
    year = "2024"
}

@article{testingbhspacetimebayesian,
    author = "Navarrete, C{\'e}sar and V{\'a}zquez-Ch{\'a}vez, Fernando and Cruz-Osorio, Alejandro and Ortiz, N{\'e}stor",
    title = "{Testing black hole space{\textendash}times with the S2 star orbit: a Bayesian comparison}",
    eprint = "2602.04980",
    archivePrefix = "arXiv",
    primaryClass = "astro-ph.HE",
    doi = "10.1093/mnras/stag059",
    journal = "Mon. Not. Roy. Astron. Soc.",
    volume = "546",
    number = "3",
    pages = "stag059",
    year = "2026"
}

@misc{Bambhaniyatesting,
    author = {Bambhaniya, Parth and Dalal, Preet and Vicentin, Giovani H. and Della Monica, Riccardo and de Gouveia Dal Pino, Elisabete M. and Patel, Bina},
    title  = {Testing the Spacetime Geometry of Sgr A* with the Relativistic Orbit of S2 star},
    year   = {2026},
    eprint = {2602.18994},
    archivePrefix = {arXiv},
    primaryClass  = {gr-qc}
}

@article{ultralightbosnicDM,
    author = "Della Monica, Riccardo and de Martino, Ivan",
    title = "{Bounding the mass of ultralight bosonic dark matter particles with the motion of the S2 star around Sgr A*}",
    eprint = "2305.10242",
    archivePrefix = "arXiv",
    primaryClass = "gr-qc",
    doi = "10.1103/PhysRevD.108.L101303",
    journal = "Phys. Rev. D",
    volume = "108",
    number = "10",
    pages = "L101303",
    year = "2023"
}

@article{narrowbosonDM,
    author = "Della Monica, Riccardo and de Martino, Ivan",
    title = "{Narrowing the allowed mass range of ultralight bosons with the S2 star}",
    eprint = "2206.03980",
    archivePrefix = "arXiv",
    primaryClass = "gr-qc",
    doi = "10.1051/0004-6361/202245150",
    journal = "Astron. Astrophys.",
    volume = "670",
    pages = "L4",
    year = "2023"
}

@article{DMspike,
    author = "Shen, Zhao-Qiang and Yuan, Guan-Wen and Jiang, Cheng-Zi and Tsai, Yue-Lin Sming and Yuan, Qiang and Fan, Yi-Zhong",
    title = "{Exploring dark matter spike distribution around the Galactic centre with stellar orbits}",
    eprint = "2303.09284",
    archivePrefix = "arXiv",
    primaryClass = "astro-ph.GA",
    doi = "10.1093/mnras/stad3282",
    journal = "Mon. Not. Roy. Astron. Soc.",
    volume = "527",
    number = "2",
    pages = "3196--3207",
    year = "2023"
}

@article{Pygropaper,
    author = "Della Monica, Riccardo",
    title = "{PyGRO: A Python integrator for General Relativistic Orbits}",
    eprint = "2504.20152",
    archivePrefix = "arXiv",
    primaryClass = "gr-qc",
    doi = "10.1051/0004-6361/202554300",
    journal = "Astron. Astrophys.",
    volume = "698",
    pages = "A193",
    year = "2025"
}

@ARTICLE{RARrecent,
       author = {{Crespi}, V. and {Arg{\"u}elles}, C.~R. and {Becerra-Vergara}, E.~A. and {Mestre}, M.~F. and {Pei{\ss}ker}, F. and {Rueda}, J.~A. and {Ruffini}, R.},
        title = "{The dynamics of S-stars and G-sources orbiting a supermassive compact object made of fermionic dark matter}",
      journal = {\mnras},
         year = 2026,
        month = feb,
       volume = {546},
       number = {1},
          eid = {staf1854},
        pages = {staf1854},
          doi = {10.1093/mnras/staf1854},
archivePrefix = {arXiv},
       eprint = {2510.19087},
 primaryClass = {astro-ph.GA},
       adsurl = {https://ui.adsabs.harvard.edu/abs/2026MNRAS.546f1854C}
}

@ARTICLE{RARold,
       author = {{Arg{\"u}elles}, C.~R. and {Mestre}, M.~F. and {Becerra-Vergara}, E.~A. and {Crespi}, V. and {Krut}, A. and {Rueda}, J.~A. and {Ruffini}, R.},
        title = "{What does lie at the Milky Way centre? Insights from the S2-star orbit precession}",
      journal = {\mnras},
         year = 2022,
        month = mar,
       volume = {511},
       number = {1},
        pages = {L35-L39},
          doi = {10.1093/mnrasl/slab126},
archivePrefix = {arXiv},
       eprint = {2109.10729},
 primaryClass = {astro-ph.GA},
       adsurl = {https://ui.adsabs.harvard.edu/abs/2022MNRAS.511L..35A}
}

@article{ZakharovRAR,
    author = "Zakharov, Alexander F.",
    title = "{Testing the Galactic Centre potential with S-stars}",
    eprint = "2108.09709",
    archivePrefix = "arXiv",
    primaryClass = "astro-ph.GA",
    doi = "10.1093/mnrasl/slab111",
    journal = {\mnras},
    volume = "513",
    number = "1",
    pages = "L6--L9",
    year = "2022"
}

@article{GalikyanNN,
  title={Neural network analysis of S-star dynamics: implications for modified gravity},
  author={N. Galikyan and Sh. Khlghatyan and Armen Kocharyan and V. G. Gurzadyan},
  journal={The European Physical Journal Plus},
  year={2023},
  volume={138},
  url={https://api.semanticscholar.org/CorpusID:263674273}
}

@misc{EyamaNN,
    author = {Eyama, Shinsei and Masada, Youhei},
    title  = {Constraining the Cosmological Constant from Stellar Orbits Around Sgr A* Using Physics-Informed Neural Networks},
    year   = {2025},
    eprint = {2508.19719},
    archivePrefix = {arXiv},
    primaryClass  = {astro-ph.CO}
}

@article{SupernovaSearchTeam1998fmf,
    author = "Riess, Adam G. and others",
    collaboration = "Supernova Search Team",
    title = "{Observational evidence from supernovae for an accelerating universe and a cosmological constant}",
    eprint = "astro-ph/9805201",
    archivePrefix = "arXiv",
    doi = "10.1086/300499",
    journal = "Astron. J.",
    volume = "116",
    pages = "1009--1038",
    year = "1998"
}

@article{SupernovaCosmologyProject1998vns,
    author = "Perlmutter, S. and others",
    collaboration = "Supernova Cosmology Project",
    title = "{Measurements of $\Omega$ and $\Lambda$ from 42 High Redshift Supernovae}",
    eprint = "astro-ph/9812133",
    archivePrefix = "arXiv",
    reportNumber = "LBNL-41801, LBL-41801",
    doi = "10.1086/307221",
    journal = "Astrophys. J.",
    volume = "517",
    pages = "565--586",
    year = "1999"
}

@article{BAO1,
    author = "Cole, Shaun and others",
    collaboration = "2dFGRS",
    title = "{The 2dF Galaxy Redshift Survey: Power-spectrum analysis of the final dataset and cosmological implications}",
    eprint = "astro-ph/0501174",
    archivePrefix = "arXiv",
    doi = "10.1111/j.1365-2966.2005.09318.x",
    journal = "Mon. Not. Roy. Astron. Soc.",
    volume = "362",
    pages = "505--534",
    year = "2005"
}

@article{BAO2,
    author = "Eisenstein, Daniel J. and others",
    collaboration = "SDSS",
    title = "{Detection of the Baryon Acoustic Peak in the Large-Scale Correlation Function of SDSS Luminous Red Galaxies}",
    eprint = "astro-ph/0501171",
    archivePrefix = "arXiv",
    reportNumber = "FERMILAB-PUB-05-057-A-CD",
    doi = "10.1086/466512",
    journal = "Astrophys. J.",
    volume = "633",
    pages = "560--574",
    year = "2005"
}

@article{BAo3,
    author = "Alam, Shadab and others",
    collaboration = "BOSS",
    title = "{The clustering of galaxies in the completed SDSS-III Baryon Oscillation Spectroscopic Survey: cosmological analysis of the DR12 galaxy sample}",
    eprint = "1607.03155",
    archivePrefix = "arXiv",
    primaryClass = "astro-ph.CO",
    doi = "10.1093/mnras/stx721",
    journal = "Mon. Not. Roy. Astron. Soc.",
    volume = "470",
    number = "3",
    pages = "2617--2652",
    year = "2017"
}

@article{Planck2018CMB,
    author = "Aghanim, N. and others",
    collaboration = "Planck",
    title = "{Planck 2018 results. VI. Cosmological parameters}",
    eprint = "1807.06209",
    archivePrefix = "arXiv",
    primaryClass = "astro-ph.CO",
    doi = "10.1051/0004-6361/201833910",
    journal = "Astron. Astrophys.",
    volume = "641",
    pages = "A6",
    year = "2020",
    note = "[Erratum: Astron.Astrophys. 652, C4 (2021)]"
}

@ARTICLE{Chernin2000,
       author = {{Chernin}, A. and {Teerikorpi}, P. and {Baryshev}, Yu.},
        title = "{Why is the Hubble flow so quiet?}",
      journal = {Advances in Space Research},
         year = 2003,
        month = jan,
       volume = {31},
       number = {2},
        pages = {459-467},
          doi = {10.1016/S0273-1177(02)00731-7},
       adsurl = {https://ui.adsabs.harvard.edu/abs/2003AdSpR..31..459C}
}

@article{Baryshev2000kw,
    author = "Baryshev, Yurij and Chernin, Arthur and Teerikorpi, Pekka",
    title = "{The Local Hubble flow: A Manifestation of dark energy}",
    journal = "{Astron. Astrophys.}",
    eprint = "astro-ph/0011528",
    archivePrefix = "arXiv",
    month = "11",
    year = "2000"
}

@article{Karachentsev:2003eh,
    author = "Karachentsev, Igor D. and Chernin, Arthur D. and Teerikorpi, Pekka",
    title = "{The Hubble Flow: Why does the cosmological expansion preserve its kinematical identity from a few Mpc distance to the observation horizon?}",
    eprint = "astro-ph/0304250",
    archivePrefix = "arXiv",
    journal = "Astrofiz.",
    volume = "46",
    pages = "399",
    year = "2003"
}

@article{Teerikorpi2005zh,
    author = "Teerikorpi, Pekka and Chernin, Arthur D. and Baryshev, Yurij V.",
    title = "{The Quiescent Hubble flow, local dark energy tests, and pairwise velocity dispersion in a Omega = 1 Universe}",
    eprint = "astro-ph/0506683",
    archivePrefix = "arXiv",
    doi = "10.1051/0004-6361:20053139",
    journal = "Astron. Astrophys.",
    volume = "440",
    pages = "791",
    year = "2005"
}

@article{Chernin2009localgroup,
  title={Dark energy and the mass of the Local Group},
  author={Arthur D. Chernin and Pekka Teerikorpi and Mauri J. Valtonen and G. Byrd and V. P. Dolgachev and L. M. Domozhilova},
  journal={arXiv: Cosmology and Nongalactic Astrophysics},
  year={2009},
  url={https://api.semanticscholar.org/CorpusID:10518259}
}

@article{Chernin2010DarkED,
  title={Dark energy domination in the Virgocentric flow},
  author={Arthur D. Chernin and Igor D. Karachentsev and Olga G. Nasonova and Pekka Teerikorpi and Mauri J. Valtonen and V. P. Dolgachev and L. M. Domozhilova and G. Byrd},
  journal={Astronomy and Astrophysics},
  year={2010},
  volume={520},
  url={https://api.semanticscholar.org/CorpusID:118464732}
}

@article{Chernin2015localflow,
    author = "Chernin, A. D. and Emelyanov, N. V. and Karachentsev, I. D.",
    title = "{Dark energy domination in the local flow of giant galaxies}",
    eprint = "1508.03485",
    archivePrefix = "arXiv",
    primaryClass = "astro-ph.CO",
    doi = "10.1093/mnras/stv144",
    journal = "Mon. Not. Roy. Astron. Soc.",
    volume = "449",
    number = "2",
    pages = "2069--2078",
    year = "2015"
}

@book{Silbergleit2019oyx,
    author = "Silbergleit, Alexander and Chernin, Arthur",
    title = "{Kepler Problem in the Presence of Dark Energy, and the Cosmic Local Flow}",
    doi = "10.1007/978-3-030-36752-7",
    publisher = "Springer",
    series = "SpringerBriefs in Physics",
    year = "2019"
}

@article{Benisty2023clf,
    author = "Benisty, David and Wagner, Jenny and Staicova, Denitsa",
    title = "{Dark energy as a critical period in binary motion: Bounds from multi-scale binaries}",
    eprint = "2310.11488",
    archivePrefix = "arXiv",
    primaryClass = "astro-ph.CO",
    doi = "10.1051/0004-6361/202348327",
    journal = "Astron. Astrophys.",
    volume = "683",
    pages = "A83",
    year = "2024"
}

@article{McLeod2016EstimatingTM,
  title={Estimating the mass of the Local Group using machine learning applied to numerical simulations},
  author={Michael McLeod and Noam I. Libeskind and Ofer Lahav and Yehuda Hoffman},
  journal={Journal of Cosmology and Astroparticle Physics},
  year={2016},
  volume={2017},
  pages={034 - 034},
  url={https://api.semanticscholar.org/CorpusID:89606320}
}

@article{LiandWhite,
    author = "Li, Yang-Shyang and White, Simon D. M.",
    title = "{Masses for the Local Group and the Milky Way}",
    eprint = "0710.3740",
    archivePrefix = "arXiv",
    primaryClass = "astro-ph",
    doi = "10.1111/j.1365-2966.2007.12748.x",
    journal = "Mon. Not. Roy. Astron. Soc.",
    volume = "384",
    pages = "1459--1468",
    year = "2008"
}

@article{Lemos:2020vhj,
    author = "Lemos, Pablo and Jeffrey, Niall and Whiteway, Lorne and Lahav, Ofer and Libeskind, Noam I. and Hoffman, Yehuda",
    title = "{Sum of the masses of the Milky Way and M31: A likelihood-free inference approach}",
    eprint = "2010.08537",
    archivePrefix = "arXiv",
    primaryClass = "astro-ph.GA",
    doi = "10.1103/PhysRevD.103.023009",
    journal = "Phys. Rev. D",
    volume = "103",
    number = "2",
    pages = "023009",
    year = "2021"
}

@BOOK{galacticdynamicsbook,
       author = {{Binney}, James and {Tremaine}, Scott},
        title = "{Galactic dynamics}",
         year = 1987,
         publisher = {Princeton University Press},
       adsurl = {https://ui.adsabs.harvard.edu/abs/1987gady.book.....B}
}

@article{Partridge2013dsa,
    author = "Partridge, Candace and Lahav, Ofer and Hoffman, Yehuda",
    title = "{Weighing the Local Group in the Presence of Dark Energy}",
    eprint = "1308.0970",
    archivePrefix = "arXiv",
    primaryClass = "astro-ph.CO",
    doi = "10.1093/mnrasl/slt109",
    journal = "Mon. Not. Roy. Astron. Soc.",
    volume = "436",
    pages = "45",
    year = "2013"
}

@article{Benistytimingargument,
    author = "Benisty, David",
    title = "{Weighing Milky Way and Andromeda in an expanding {\ensuremath{\Lambda}}CDM Universe - Decreasing the Local Group mass}",
    eprint = "2401.09546",
    archivePrefix = "arXiv",
    primaryClass = "astro-ph.CO",
    doi = "10.1051/0004-6361/202449884",
    journal = "Astron. Astrophys.",
    volume = "689",
    pages = "L1",
    year = "2024"
}

@article{Benisty2024galaxygroups,
    author = "Benisty, David and Chaichian, Moshe M. and Tureanu, Anca",
    title = "{Galaxy groups in the presence of cosmological constant: Increasing the masses of groups}",
    eprint = "2405.14944",
    archivePrefix = "arXiv",
    primaryClass = "astro-ph.GA",
    doi = "10.1016/j.physletb.2024.139033",
    journal = "Phys. Lett. B",
    volume = "858",
    pages = "139033",
    year = "2024"
}

@article{Benistyrotationcurve,
    author = {Benisty, David and Vasak, David and Struckmeier, J{\"u}rgen and Stoecker, Horst},
    title = "{Bounding the cosmological constant using galactic rotation curves from SPARC dataset}",
    eprint = "2405.16650",
    archivePrefix = "arXiv",
    primaryClass = "astro-ph.CO",
    doi = "10.1103/PhysRevD.110.063028",
    journal = "Phys. Rev. D",
    volume = "110",
    number = "6",
    pages = "063028",
    year = "2024"
}

@article{Benisty:2023vbz,
    author = "Benisty, David and Davis, Anne-Christine and Evans, N. Wyn",
    title = "{Constraining Dark Energy from the Local Group Dynamics}",
    eprint = "2306.14963",
    archivePrefix = "arXiv",
    primaryClass = "astro-ph.CO",
    doi = "10.3847/2041-8213/ace90b",
    journal = "Astrophys. J. Lett.",
    volume = "953",
    number = "1",
    pages = "L2",
    year = "2023"
}

@article{Benisty:2021davis,
    author = "Benisty, David and Davis, Anne-Christine",
    title = "{Dark energy interactions near the Galactic Center}",
    eprint = "2108.06286",
    archivePrefix = "arXiv",
    primaryClass = "astro-ph.CO",
    doi = "10.1103/PhysRevD.105.024052",
    journal = "Phys. Rev. D",
    volume = "105",
    number = "2",
    pages = "024052",
    year = "2022"
}

@misc{ZakharovAdS,
    author = {Zakharov, Alexander F.},
    title  = {Are signatures of anti-de-Sitter black hole at the Galactic Center?},
    year   = {2014},
    eprint = {1407.2591},
    archivePrefix = {arXiv},
    primaryClass  = {astro-ph.GA}
}

@article{Prado-Fuentes2025nvl,
    author = "Prado-Fuentes, Reginaldo and Aros, Rodrigo and Estrada, Milko and Astudillo, Bastian",
    title = "{Constructing Regular Lovelock Black Holes with Degenerate Vacuum and {\ensuremath{\Lambda}} {\ensuremath{<}} 0 Using the Gravitational Tension{\textemdash}Shadow Analysis}",
    eprint = "2502.07992",
    archivePrefix = "arXiv",
    primaryClass = "gr-qc",
    doi = "10.3390/universe11100338",
    journal = "Universe",
    volume = "11",
    number = "10",
    pages = "338",
    year = "2025"
}

@article{Perlick:2021aok,
    author = "Perlick, Volker and Tsupko, Oleg Yu.",
    title = "{Calculating black hole shadows: Review of analytical studies}",
    eprint = "2105.07101",
    archivePrefix = "arXiv",
    primaryClass = "gr-qc",
    doi = "10.1016/j.physrep.2021.10.004",
    journal = "Phys. Rept.",
    volume = "947",
    pages = "1--39",
    year = "2022"
}

@article{Roy:2020dyy,
    author = "Roy, Rittick and Chakrabarti, Sayan",
    title = "{Study on black hole shadows in asymptotically de Sitter spacetimes}",
    eprint = "2003.14107",
    archivePrefix = "arXiv",
    primaryClass = "gr-qc",
    doi = "10.1103/PhysRevD.102.024059",
    journal = "Phys. Rev. D",
    volume = "102",
    number = "2",
    pages = "024059",
    year = "2020"
}

@article{Firouzjaee:2019aij,
    author = "Firouzjaee, Javad T. and Allahyari, Alireza",
    title = "{Black hole shadow with a cosmological constant for cosmological observers}",
    eprint = "1905.07378",
    archivePrefix = "arXiv",
    primaryClass = "astro-ph.CO",
    doi = "10.1140/epjc/s10052-019-7464-2",
    journal = "Eur. Phys. J. C",
    volume = "79",
    number = "11",
    pages = "930",
    year = "2019"
}

@article{Keckredshift,
    author = "Do, Tuan and others",
    title = "{Relativistic redshift of the star S0-2 orbiting the Galactic center supermassive black hole}",
    eprint = "1907.10731",
    archivePrefix = "arXiv",
    primaryClass = "astro-ph.GA",
    doi = "10.1126/science.aav8137",
    journal = "Science",
    volume = "365",
    number = "6454",
    pages = "664--668",
    year = "2019"
}

@article{Kottler1918cxc,
    author = "Kottler, Friedrich",
    title = {{{\"U}ber die physikalischen Grundlagen der Einsteinschen Gravitationstheorie}},
    doi = "10.1002/andp.19183611402",
    journal = "Annalen Phys.",
    volume = "361",
    number = "14",
    pages = "401--462",
    year = "1918"
}

@article{Perlick:2018iye,
    author = "Perlick, Volker and Tsupko, Oleg Yu. and Bisnovatyi-Kogan, Gennady S.",
    title = "{Black hole shadow in an expanding universe with a cosmological constant}",
    eprint = "1804.04898",
    archivePrefix = "arXiv",
    primaryClass = "gr-qc",
    doi = "10.1103/PhysRevD.97.104062",
    journal = "Phys. Rev. D",
    volume = "97",
    number = "10",
    pages = "104062",
    year = "2018"
}

@book{hairer1993ode1,
  title     = {Solving Ordinary Differential Equations I: Nonstiff Problems},
  author    = {Hairer, Ernst and N{\o}rsett, Syvert Paul and Wanner, Gerhard},
  series    = {Springer Series in Computational Mathematics},
  volume    = {8},
  edition   = {2},
  year      = {1993},
  publisher = {Springer-Verlag},
  address   = {Berlin, Heidelberg}
}

@article{catalogue,
  author={Stefan Gillessen and Philipp M. Plewa and Frank Eisenhauer and R. Sari and Idel Waisberg and Maryam Habibi and Oliver Pfuhl and E. M. George and Jason Dexter and S. D. von Fellenberg and Thomas Ott and Reinhard Genzel},
  title = {25yrs monitoring of stellar orbits in the GC}, 
  journal = {Centre de Donnees astronomique de Strasbourg (CDS)},
year= {2016},
  doi = {https://doi.org/10.26093/cds/vizier.18370030}
}

@ARTICLE{Plewaoffset,
       author = {{Plewa}, P.~M. and {Gillessen}, S. and {Eisenhauer}, F. and {Ott}, T. and {Pfuhl}, O. and {George}, E. and {Dexter}, J. and {Habibi}, M. and {Genzel}, R. and {Reid}, M.~J. and {Menten}, K.~M.},
        title = "{Pinpointing the near-infrared location of Sgr A* by correcting optical distortion in the NACO imager}",
      journal = {\mnras},
         year = 2015,
        month = nov,
       volume = {453},
       number = {3},
        pages = {3234-3244},
          doi = {10.1093/mnras/stv1910},
archivePrefix = {arXiv},
       eprint = {1509.01941},
 primaryClass = {astro-ph.GA},
       adsurl = {https://ui.adsabs.harvard.edu/abs/2015MNRAS.453.3234P}
}

@article{Riess2021SH0ES,
    author = "Riess, Adam G. and others",
    title = "{A Comprehensive Measurement of the Local Value of the Hubble Constant with 1 km s$^{-1}$ Mpc$^{-1}$ Uncertainty from the Hubble Space Telescope and the SH0ES Team}",
    eprint = "2112.04510",
    archivePrefix = "arXiv",
    primaryClass = "astro-ph.CO",
    doi = "10.3847/2041-8213/ac5c5b",
    journal = "Astrophys. J. Lett.",
    volume = "934",
    number = "1",
    pages = "L7",
    year = "2022"
}

@article{Pantheonplus,
    author = "Brout, Dillon and others",
    title = "{The Pantheon+ Analysis: Cosmological Constraints}",
    eprint = "2202.04077",
    archivePrefix = "arXiv",
    primaryClass = "astro-ph.CO",
    doi = "10.3847/1538-4357/ac8e04",
    journal = "Astrophys. J.",
    volume = "938",
    number = "2",
    pages = "110",
    year = "2022"
}

@article{emcee,
  title={emcee: The MCMC Hammer},
  author={Daniel Foreman-Mackey and David W. Hogg and Dustin Lang and Jonathan B. Goodman},
  journal={Publications of the Astronomical Society of the Pacific},
  year={2012},
  volume={125},
  pages={306 - 312},
  url={https://api.semanticscholar.org/CorpusID:88518555}
}

@article{optimization,
    author = "Richard H., Byrd and Peihuang, Lu and Jorge, Nocedal and Ciyou, Zhu",
    title = "{A Limited Memory Algorithm for Bound Constrained Optimization}",
    doi = "10.1137/0916069",
    journal = "SIAM J. Sci. Comput.",
    volume = "16",
    number = "5",
    pages = "1190--1208",
    year = "2006"
}

@ARTICLE{autocorrelation,
       author = {{Goodman}, Jonathan and {Weare}, Jonathan},
        title = "{Ensemble samplers with affine invariance}",
      journal = {Communications in Applied Mathematics and Computational Science},
         year = 2010,
        month = jan,
       volume = {5},
       number = {1},
        pages = {65-80},
          doi = {10.2140/camcos.2010.5.65},
       adsurl = {https://ui.adsabs.harvard.edu/abs/2010CAMCS...5...65G}
}

@ARTICLE{sdsasdsshadow1,
       author = {{Stuchl{\'\i}k}, Z. and {Hled{\'\i}k}, S.},
        title = "{Some properties of the Schwarzschild-de Sitter and Schwarzschild-anti-de Sitter spacetimes}",
      journal = {\prd},
         year = 1999,
        month = aug,
       volume = {60},
       number = {4},
          eid = {044006},
        pages = {044006},
          doi = {10.1103/PhysRevD.60.044006},
       adsurl = {https://ui.adsabs.harvard.edu/abs/1999PhRvD..60d4006S}
}

@article{sdsshadownotes,
    author = "Bisnovatyi-Kogan, G. S. and Tsupko, O. Yu. and Perlick, V.",
    title = "{Shadow of a black hole at local and cosmological distances}",
    eprint = "1910.10514",
    archivePrefix = "arXiv",
    primaryClass = "gr-qc",
    doi = "10.22323/1.362.0009",
    journal = "PoS",
    volume = "MULTIF2019",
    pages = "009",
    year = "2019"
}

@BOOK{rindlerbook,
       author = {{Rindler}, Wolfgang},
        title = "{Relativity: special, general, and cosmological}",
         year = 2001,
         publisher = "Oxford University Press",
       adsurl = {https://ui.adsabs.harvard.edu/abs/2001rsgc.book.....R}
}

@article{Kerrsdsprecession,
    author = "Kerr, Andrew W. and Hauck, John C. and Mashhoon, Bahram",
    title = "{Standard clocks, orbital precession and the cosmological constant}",
    eprint = "gr-qc/0301057",
    archivePrefix = "arXiv",
    doi = "10.1088/0264-9381/20/13/320",
    journal = "Class. Quant. Grav.",
    volume = "20",
    pages = "2727",
    year = "2003"
}

@article{Iorio:2007ub,
    author = "Iorio, Lorenzo",
    title = "{Solar System motions and the cosmological constant: A New approach}",
    eprint = "0710.2610",
    archivePrefix = "arXiv",
    primaryClass = "gr-qc",
    doi = "10.1155/2008/268647",
    journal = "Adv. Astron.",
    volume = "2012",
    pages = "268647",
    year = "2012",
    note = "[Erratum: Adv.Astron. 2012, 507892 (2012)]"
}

@article{Arakida:2012ya,
    author = "Arakida, Hideyoshi",
    title = "{Note on the perihelion/periastron advance due to cosmological constant}",
    eprint = "1212.6289",
    archivePrefix = "arXiv",
    primaryClass = "gr-qc",
    doi = "10.1007/s10773-012-1458-2",
    journal = "Int. J. Theor. Phys.",
    volume = "52",
    pages = "1408--1414",
    year = "2013"
}

@article{Ruggiero:2010yn,
    author = "Ruggiero, Matteo Luca",
    title = "{Perturbations of Keplerian Orbits in Stationary Spherically Symmetric Spacetimes}",
    eprint = "1010.2114",
    archivePrefix = "arXiv",
    primaryClass = "gr-qc",
    doi = "10.1142/S0218271814500497",
    journal = "Int. J. Mod. Phys. D",
    volume = "23",
    pages = "1450049",
    year = "2014"
}

@article{DESI1,
    author = "Adame, A. G. and others",
    collaboration = "DESI",
    title = "{DESI 2024 VII: cosmological constraints from the full-shape modeling of clustering measurements}",
    eprint = "2411.12022",
    archivePrefix = "arXiv",
    primaryClass = "astro-ph.CO",
    reportNumber = "FERMILAB-PUB-24-0854-PPD",
    doi = "10.1088/1475-7516/2025/07/028",
    journal = "JCAP",
    volume = "07",
    pages = "028",
    year = "2025"
}

@article{DESI2,
    author = "Calderon, R. and others",
    collaboration = "DESI",
    title = "{DESI 2024: reconstructing dark energy using crossing statistics with DESI DR1 BAO data}",
    eprint = "2405.04216",
    archivePrefix = "arXiv",
    primaryClass = "astro-ph.CO",
    doi = "10.1088/1475-7516/2024/10/048",
    journal = "JCAP",
    volume = "10",
    pages = "048",
    year = "2024"
}

@article{DESI3DR2,
    author = "Lodha, K. and others",
    collaboration = "DESI",
    title = "{Extended dark energy analysis using DESI DR2 BAO measurements}",
    eprint = "2503.14743",
    archivePrefix = "arXiv",
    primaryClass = "astro-ph.CO",
    reportNumber = "FERMILAB-PUB-25-0164-PPD",
    doi = "10.1103/w4c6-1r5j",
    journal = "Phys. Rev. D",
    volume = "112",
    number = "8",
    pages = "083511",
    year = "2025"
}

@article{DESIDR2,
    author = "Abdul Karim, M. and others",
    collaboration = "DESI",
    title = "{DESI DR2 results. II. Measurements of baryon acoustic oscillations and cosmological constraints}",
    eprint = "2503.14738",
    archivePrefix = "arXiv",
    primaryClass = "astro-ph.CO",
    reportNumber = "FERMILAB-PUB-25-0169-PPD",
    doi = "10.1103/tr6y-kpc6",
    journal = "Phys. Rev. D",
    volume = "112",
    number = "8",
    pages = "083515",
    year = "2025"
}

@article{Shaymatov2023jfa,
    author = {Shaymatov, Sanjar and Ahmedov, Bobomurat and De Laurentis, Mariafelicia and Jamil, Mubasher and Wu, Qiang and Wang, Anzhong and Azreg-A{\"\i}nou, Mustapha},
    title = "{On the Parameters of the Spherically Symmetric Parameterized Rezzolla{\textendash}Zhidenko Spacetime through Solar System Tests, the Orbit of the S2 Star about Sgr A*, and Quasiperiodic Oscillations}",
    eprint = "2307.10804",
    archivePrefix = "arXiv",
    primaryClass = "gr-qc",
    doi = "10.3847/1538-4357/acfcba",
    journal = "Astrophys. J.",
    volume = "959",
    number = "1",
    pages = "6",
    year = "2023"
}

@article{Grould2017bsw,
    author = "Grould, M. and Vincent, F. H. and Paumard, T. and Perrin, G.",
    title = "{General relativistic effects on the orbit of the S2 star with GRAVITY}",
    eprint = "1709.04492",
    archivePrefix = "arXiv",
    primaryClass = "astro-ph.HE",
    doi = "10.1051/0004-6361/201731148",
    journal = "Astron. Astrophys.",
    volume = "608",
    pages = "A60",
    year = "2017"
}

@book{doublestarsbook,
  title     = {Double stars},
  author    = {{Heintz}, W.~D.},
  series    = {GEOPHYSICS AND ASTROPHYSICS MONOGRAPHS, AN INTERNATIONAL SERIES OF FUNDAMENTAL TEXTBOOKS},
  volume    = {15},
  year      = {1978},
  publisher = {D. Reidel Publishing Company},
  address   = {DORDRECHT, HOLLAND / BOSTON, U.S.A/ LONDON, ENGLAND}
}

@ARTICLE{NTT,
       author = {{Hofmann}, R. and {Eckart}, A. and {Genzel}, R. and {Drapatz}, S.},
        title = "{High Resolution K-Band Images of the Galactic Centre}",
      journal = {\apss},
         year = 1993,
        month = jul,
       volume = {205},
       number = {1},
        pages = {1-4},
          doi = {10.1007/BF00657949},
       adsurl = {https://ui.adsabs.harvard.edu/abs/1993Ap&SS.205....1H}
}

@INPROCEEDINGS{NACO,
       author = {{Lenzen}, Rainer and {Hofmann}, Reiner and {Bizenberger}, Peter and {Tusche}, Andreas},
        title = "{CONICA: the high-resolution near-infrared camera for the ESO VLT}",
    booktitle = {Infrared Astronomical Instrumentation},
         year = 1998,
       editor = {{Fowler}, Albert M.},
       series = {Society of Photo-Optical Instrumentation Engineers (SPIE) Conference Series},
       volume = {3354},
        month = aug,
        pages = {606-614},
          doi = {10.1117/12.317287},
       adsurl = {https://ui.adsabs.harvard.edu/abs/1998SPIE.3354..606L}
}

@INPROCEEDINGS{NACO2,
       author = {{Rousset}, Gerard and {Lacombe}, Francois and {Puget}, Pascal and {Hubin}, Norbert N. and {Gendron}, Eric and {Conan}, Jean-Marc and {Kern}, Pierre Y. and {Madec}, Pierre-Yves and {Rabaud}, Didier and {Mouillet}, David and {Lagrange}, Anne-Marie and {Rigaut}, Francois J.},
        title = "{Design of the Nasmyth adaptive optics system (NAOS) of the VLT}",
    booktitle = {Adaptive Optical System Technologies},
         year = 1998,
       editor = {{Bonaccini}, Domenico and {Tyson}, Robert K.},
       series = {Society of Photo-Optical Instrumentation Engineers (SPIE) Conference Series},
       volume = {3353},
        month = sep,
        pages = {508-516},
          doi = {10.1117/12.321686},
       adsurl = {https://ui.adsabs.harvard.edu/abs/1998SPIE.3353..508R}
}

@article{velocity1,
    author = "Eisenhauer, F. and Schoedel, R. and Genzel, R. and Ott, T. and Tecza, M. and Abuter, R. and Eckart, A. and Alexander, T.",
    title = "{A geometric determination of the distance to the galactic center}",
    eprint = "astro-ph/0306220",
    archivePrefix = "arXiv",
    doi = "10.1086/380188",
    journal = "Astrophys. J. Lett.",
    volume = "597",
    pages = "L121--L124",
    year = "2003"
}

@article{velocity2,
    author = "Eisenhauer, Frank and others",
    title = "{Sinfoni - integral field spectroscopy at 50 milli-arcsecond resolution with the eso vlt}",
    eprint = "astro-ph/0306191",
    archivePrefix = "arXiv",
    doi = "10.1117/12.459468",
    journal = "Proc. SPIE Int. Soc. Opt. Eng.",
    volume = "4841",
    pages = "1548--1561",
    year = "2003"
}

@ARTICLE{velocity3,
       author = {{Bonnet}, Henri and {Abuter}, Robert and {Baker}, Andrew and {Bornemann}, Walter and {Brown}, Anthony and {Castillo}, Roberto and {Conzelmann}, Ralf and {Damster}, Romuald and {Davies}, Richard and {Delabre}, Bernard and {Donaldson}, Rob and {Dumas}, Christophe and {Eisenhauer}, Frank and {Elswijk}, Eddie and {Fedrigo}, Enrico and {Finger}, Gert and {Gemperlein}, Hans and {Genzel}, Reinhard and {Gilbert}, Andrea and {Gillet}, Gordon and {Goldbrunner}, Armin and {Horrobin}, Matthew and {Ter Horst}, Rik and {Huber}, Stefan and {Hubin}, Norbert and {Iserlohe}, Christof and {Kaufer}, Andreas and {Kissler-Patig}, Markus and {Kragt}, Jan and {Kroes}, Gabby and {Lehnert}, Matthew and {Lieb}, Werner and {Liske}, Jochen and {Lizon}, Jean-Louis and {Lutz}, Dieter and {Modigliani}, Andrea and {Monnet}, Guy and {Nesvadba}, Nicole and {Patig}, Jona and {Pragt}, Johan and {Reunanen}, Juha and {R{\"o}hrle}, Claudia and {Rossi}, Silvio and {Schmutzer}, Riccardo and {Schoenmaker}, Ton and {Schreiber}, J{\"u}rgen and {Stroebele}, Stefan and {Szeifert}, Thomas and {Tacconi}, Linda and {Tecza}, Matthias and {Thatte}, Niranjan and {Tordo}, Sebastien and {van der Werf}, Paul and {Weisz}, Harald},
        title = "{First light of SINFONI at the VLT}",
      journal = {The Messenger},
         year = 2004,
        month = sep,
       volume = {117},
        pages = {17-24},
       adsurl = {https://ui.adsabs.harvard.edu/abs/2004Msngr.117...17B}
}
\end{document}